\documentclass[twocolumn,aps,nofootinbib]{revtex4}
\usepackage[pdftex]{graphicx}
\usepackage{epstopdf}
\usepackage{latexsym}
\usepackage{amsmath}
\usepackage{amssymb}
\usepackage{color}
\usepackage{mathrsfs}
\usepackage{lipsum}

\usepackage[center]{subfigure}

\begin{document}

 \newcommand{\bq}{\begin{equation}}
 \newcommand{\eq}{\end{equation}}
 \newcommand{\bqn}{\begin{eqnarray}}
 \newcommand{\eqn}{\end{eqnarray}}
 \newcommand{\nb}{\nonumber}
 \newcommand{\lb}{\label}
 \newcommand{\be}{\begin{equation}}
\newcommand{\en}{\end{equation}}
\newcommand{\PRL}{Phys. Rev. Lett.}
\newcommand{\PL}{Phys. Lett.}
\newcommand{\PR}{Phys. Rev.}
\newcommand{\CQG}{Class. Quantum Grav.}

\title{Echoes of bimodal axial gravitational perturbations in a uniform-density star in Einstein-{\AE}ther gravity}

\author{Kai Lin}
\email{kailin@if.usp.br}
\affiliation{Universidade Federal de Campina Grande, Campina Grande, PB, Brasil}
\affiliation{Instituto de F\'isica, Universidade de S\~ao Paulo, S\~ao Paulo, Brasil}

\author{Wei-Liang Qian}
\email{wlqian@usp.br (corresponding author)}
\affiliation{Escola de Engenharia de Lorena, Universidade de S\~ao Paulo, 12602-810, Lorena, SP, Brazil}
\affiliation{Faculdade de Engenharia de Guaratinguet\'a, Universidade Estadual Paulista, 12516-410, Guaratinguet\'a, SP, Brazil}
\affiliation{Center for Gravitation and Cosmology, School of Physical Science and Technology, Yangzhou University, 225002, Yangzhou, Jiangsu, China}

\author{Alan B. Pavan}
\email{alan@unifei.edu.br}
\affiliation{Universidade Federal de Itajub\'a, Instituto de F\'isica e Qu\'imica, Itajub\'a, MG, Brasil}

\author{Amilcar Rabelo de Queiroz}
\email{amilcarq@df.ufcg.edu.br}
\affiliation{Universidade Federal de Campina Grande, Campina Grande, PB, Brasil}

\author{Elcio Abdalla}
\email{eabdalla@usp.br}
\affiliation{Instituto de F\'isica, Universidade de S\~ao Paulo, S\~ao Paulo, Brasil}

%


\begin{abstract}
This paper studies axial gravitational perturbations of a uniform-density star in scalar-Einstein-{\AE}ther theory.
By applying the Israel junction conditions explicitly in the presence of a scalar field minimally coupled to the gravitational sector, it is shown that a nontrivial scalar profile cannot be sustained, as it induces a divergence at the stellar center.
Since analytical solutions are unattainable, the background metric is determined through numerical integration for a few representative configurations.
For axial gravitational perturbations, it is found that the system of equations of motion cannot be decoupled as long as the {\AE}ther parameter $c_i$ does not vanish.
Subsequently, the dynamics of the system can be simplified to two coupled equations that describe vector and tensor perturbations with distinct wave velocities $c_V$ and $c_T$, giving rise to a bimodal system.
It is shown that as the stellar radius is smaller than that of the maximum of the vacuum Schwarzschild-type effective potential, a potential well is formed, leading to the emergence of echo phenomenon for the axial gravitational perturbations.
When the stellar radius exceeds the {\it could-have-been} maximum, the resulting effective potential decreases monotonically, and the wave propagation is primarily dictated by the discontinuity occurring at the star's surface, producing a type of more attenuated echo waves.
In addition, we explore the specific properties of the resultant bimodal medium consisting of two degrees of freedom with distinct sound speeds.
However, it is understood that such a characteristic does not lead to observational implications, and subsequently hardly offers a potential empirical means to constrain specific metric parameters of the Einstein-{\AE}ther.
We present numerical calculations and discuss the potential implications of our findings.
\end{abstract}


\maketitle

\section{Introduction}
\renewcommand{\theequation}{1.\arabic{equation}} 
\setcounter{equation}{0}

General relativity is undoubtedly one of the most prominent physics achievements of the last century. 
In the past decade, with the discovery of gravitational waves and the unveiling of black hole images, nearly all of the predictions made by general relativity have been confirmed~\cite{LIGO1, LIGO2, LIGO3, LIGO4, LIGO5, LIGO6, LIGO7, LIGO8, LIGO9, BHimage1, BHimage2, BHimage3, BHimage4, BHimage5, BHimage6, BHimage7}. 
Nevertheless, general relativity still has its shortcomings: firstly, the dark energy and dark matter discovered in the universe have long remained inadequately explained, becoming the biggest unsolved puzzles in astronomy and cosmology; additionally, in theoretical physics, it has been found that general relativity is a non-renormalizable theory, which means that Einstein's theory is valid only at energies much smaller than the Planck scale.

Various modified gravity theories have been proposed to address these challenges.
Among these, the Einstein-{\AE}ther theory is a particularly notable candidate~\cite{AetherEinstein}.
Post-Newtonian approximation calculations indicate that, with appropriate parameter adjustments, the Einstein-{\AE}ther theory can replicate results identical to those of general relativity, satisfying all post-Newtonian tests validating Einstein's theory~\cite{postNewton1,postNewton2}.
Additionally, the theory converges with the Horava-Lifshitz theory at low energies, positioning it as a promising candidate for quantum gravity~\cite{HLtoEA1, HLtoEA2, HLtoEA3}.
A vital aspect of the theory is the assumption of Lorentz-symmetry breaking, achieved by introducing a timelike vector field, referred to as the ``{\ae}ther'', into the action.
Consequently, in this theory, the propagation speeds of the scalar ($c_S$), vector ($c_V$), and tensor ($c_T$) degrees of freedom, determined by four coupling constants, may differ from the speed of light~\cite{EAspeed1, EAspeed2}.
This leads to challenges in defining the black hole's event horizon.
The event horizon, traditionally defined with respect to the speed of light, might be crossed by superluminal particles predicted by the theory.
Nevertheless, studies have shown the existence of a Universal Horizon within the black hole's event horizon, which prevents even infinitely fast particles from escaping~\cite{UniveralHorizon1, UniveralHorizon2, UniveralHorizon3}.
Despite this, the standard coordinates $(t, r, \theta, \varphi)$ are inadequate for describing black hole spacetimes in this theory.
These coordinates remain effective for spacetimes without black hole horizons, such as cosmological or wormhole spacetimes.

This study investigates axial gravitational perturbations in uniform-density stars as spacetime metric solutions within the scalar-Einstein-{\AE}ther theory.
Sec.~II derives the spacetime metric for a static, uniform-density star.
As the minimally coupled scalar field does not possess a nontrivial solution, one obtains a spherically symmetric vacuum solution for the star's exterior, analogous to the Schwarzschild metric.
However, obtaining an analytical solution for the interior metric is generally challenging.
Instead, we present numerical solutions derived using the finite difference method (FDM).
The background metric of static, uniform-density stars depends solely on $c_{14}\equiv c_1+c_4$ within the Einstein-{\AE}ther theory.
Given this background metric, we analyze axial gravitational perturbations and derive the corresponding equations of motion.
It is shown that the wave equation for this perturbation cannot be fully decoupled when $c_{14} \neq 0$.
At best, one can obtain a coupled system of vector and tensor wave equations, with the field dynamics governed entirely by three parameters: $c_{14}$, $c_V$, and $c_T$.
In Sec.~III, we explore the temporal evolution of axial gravitational perturbations.
We demonstrate the existence of echoes~\cite{echoPani1,echoPani2,echoPani3,echoPani4}.
Specifically, when the star's radius is smaller than the maximum of the Schwarzschild-type effective potential, a potential well is formed inside the star, leading to echoes.
Conversely, when the radius exceeds this maximum, the effective potential lacks a local maximum and decreases monotonically with the radial coordinate.
The latter scenario produces a distinct type of echoes~\cite{echoTwo1,echoTwo2}, driven by the discontinuity at the star's surface.
Numerically, the waveform of the echoes exhibits more significant attenuation.
Finally, we explore the potential implications of the bimodal nature of the model, arising from the distinct propagation speeds of the vector and tensor degrees of freedom.
However, it can be argued that the very construction of the theory implies that this feature could hardly be used to constrain the metric parameters of the underlying theory via experimental observation.
Sec.~IV presents the concluding remarks.

\section{Uniform density star solution in scalar-Einstein-{\AE}ther Gravity}
\renewcommand{\theequation}{2.\arabic{equation}} \setcounter{equation}{0}
The action of the Einstein-{\AE}ther theory~\cite{HLtoEA1,HLtoEA2,HLtoEA3,EAG,EAspeed1,EAspeed2,UniveralHorizon1,UniveralHorizon2,UniveralHorizon3,wormhole} when minimally coupled with a scalar field $\Psi$, is given by
\bqn
\lb{Action1}
S&=&\frac{1}{16\pi G_\mathrm{\ae}}\int\sqrt{-g}d^4x\left[R
-2g^{\alpha\beta}D_\alpha\Psi D_\beta\Psi\right.\nb\\
&&-\left(c_1g^{\alpha\beta}g_{\mu\nu}+c_2\delta^\alpha_\mu\delta^\beta_\nu+c_3\delta^\alpha_\nu\delta^\beta_\mu-c_4u^\alpha u^\beta g_{\mu\nu}\right)\nb\\
&&\left.\times\left(D_\alpha u^\mu\right)\left(D_\beta u^\nu\right)+\lambda\left(u^\rho u_\rho+1\right)\right]+S_m,
\eqn
where $S_m$ is the action of the remaining matter field, the constant $G_\mathrm{\ae}$ is related to the Newtonian counterpart $G_N$ by a $c_i$-dependent rescaling $G_\mathrm{\ae}\equiv\left(1-c_{14}/2\right)G_N$~\cite{EAG}, and we write $c_{14}\equiv c_1+c_4$, $c_\pm\equiv c_1\pm c_3$ and $c_{123}=c_1+c_2+c_3$. 
Also, we set $\kappa_m=8\pi G_\mathrm{\ae}=c=1$.

The resulting field equations are
\bqn
\lb{FieldEqu1}
R_{\mu\nu}-\frac{1}{2}g_{\mu\nu}R-S_{\mu\nu}&=&T_{\mu\nu} ,\nb\\
\text{\AE}_\mu&=&0 ,\nb\\
g^{\alpha\beta}D_\alpha D_\beta\Psi&=&0 ,\nb\\
g_{\alpha\beta}u^\alpha u^\beta&=&-1 ,
\eqn
where
\bqn
\lb{FieldEqu2}
S_{\alpha\beta}&\equiv&D_{\mu}\left[J^\mu_{(\alpha}u_{\beta)}+J_{(\alpha\beta)}u^\mu-u_{(\beta}J_{\alpha)}^\mu\right]\nb\\
&&+c_1\left[\left(D_\alpha u_\mu\right)\left(D_\beta u^\mu\right)-\left(D_\mu u_\alpha\right)\left(D^\mu u_\beta\right)\right]\nb\\
&&+c_4a_\alpha a_\beta+\lambda u_\alpha u_\beta-\frac{1}{2}g_{\alpha\beta}J^\mu_\nu D_\mu u^\nu ,\nb\\
\text{\AE}_\mu&\equiv&D_\nu J^\nu_\mu+c_4 a_\nu D_\mu u^\nu+\lambda u_\mu,\nb\\
T_{\mu\nu}&\equiv&(\rho+P)U_\mu U_\nu+Pg_{\mu\nu}\nb\\
&&+2\left(D_\mu\Psi D_\nu\Psi-\frac{1}{2}g_{\mu\nu}D^\alpha\Psi D_\alpha \Psi\right) ,\nb\\
\eqn
with
\bqn
\lb{FieldEqu3}
J^\alpha_\mu&\equiv&\left(c_1g^{\alpha\beta}g_{\mu\nu}+c_2\delta^\alpha_\mu\delta^\beta_\nu+c_3\delta^\alpha_\nu\delta^\beta_\mu\right.\nb\\
&&\left.-c_4u^\alpha u^\beta g_{\mu\nu}\right)D_{\beta}u^\nu ,\nb\\
a^\mu&\equiv&u^\alpha D_\alpha u^\mu ,\nb\\
U^\mu&=&\delta^\mu_t=(1,0,0,0) ,
\eqn
where the matter field is characterized by the pressure $P$ and density $\rho$.
From the above field equations, we have
\bqn
\lb{FieldEqu4}
\lambda=u_\beta D_\alpha J^{\alpha\beta}+c_4a_\rho a^\rho .
\eqn

For a static star with spherical symmetry, one considers the following ansatz
\bqn
\lb{metric1}
ds^2&=&-h(r)dt^2+\frac{dr^2}{f(r)}+r^2\left(d\theta^2+\sin^2\theta d\varphi^2\right),\nb\\
u_\mu&=&\sqrt{h(r)}\delta^t_\mu ,\nb\\
\Psi&=&\Psi(r) ,
\eqn
which corresponds to the case by choosing $\alpha(r)=0$ in a more general form~\cite{Hsu:2024ftc}
\bqn
\left\lbrack u_{\mu} \right\rbrack = \left\lbrack\sqrt{h(r)}\cosh(\alpha(r)),\frac{1}{\sqrt{f(r)}}\sinh(\alpha(r)),0,0 \right\rbrack .\nb
\eqn
However, as discussed in Appx.~\ref{appA}, such a star metric does not accommodate a nontrivial scalar field, owing to the second Israel junction condition.
By suppressing the scalar field, the field equations are simplified to:
\bqn
\lb{Equation1}
&&f'+f\left(\frac{2}{r}+\frac{h'}{h}\right)-\frac{2}{r}\nb\\
&&~~~~+\frac{2r}{2-c_{14}}\left[\rho+(2c_{14}-1)P\right]=0 ,\nb\\
&&h'+\frac{4}{c_{14}}\frac{h}{r}\nb\\
&&~~~~-\frac{2\sqrt{2}h}{c_{14}r\sqrt{f}}\sqrt{c_{14}+(2-c_{14})f+c_{14}r^2P}=0 ,\nb\\
&&P'+\frac{h'}{2h}\left(P+\rho\right)=0 .
\eqn

For the exterior of the star $r>r_s$, we have $\rho=P=0$.
For both the interior and exterior of the star, analytical solutions are generally difficult to obtain.
Nonetheless, numerical methods can be used. 
The interior solution and the vacuum solution are connected at the star's surface by Israel's first junction conditions~\cite{agr-collapse-thin-shell-03, book-general-relativity-Poisson}: 
\bqn
f_\text{inside}(r_s) &=& f_\text{outside}(r_s), \nb\\
h_\text{inside}(r_s) &=& h_\text{outside}(r_s), \label{bdCon}
\eqn
together with a vanishing pressure $P(r_s)=0$.
Observing the second line of Eqs.~\eqref{Equation1}, one notes that the continuation of $f$, $h$, and $P$ on the star's surface implies that $h'$ is continuous across the surface.
However, the jump in $\rho$ leads to the discontinuity in $f'$ at $r=r_s$, which implies a nonvanishing energy-momentum tensor on the surface.
Further discussions about the above junction condition can be found in Appx.~\ref{appA}.
It is noting that the last line of Eqs.~\eqref{Equation1} can be integrated once, starting from the surface of the star and moving inward or outward, to give $P(r) = \rho(r)\left( \sqrt{\frac{h\left( r_{s} \right)}{h(r)}} - 1 \right) $.

For given values of $\rho$ and $c_{14}$, we can numerically integrate inward and outward by Eqs.~\eqref{Equation1} from the star's surface, where the junction conditions Eqs.~\eqref{bdCon} are satisfied.
A valid numerical solution is obtained by tuning $f(r_s)$ and $h(r_s)$, satisfying the requirement that the spacetime is asymptotically flat.
The numerical results of $f=f(r)$, $h=h(r)$, and $P=P(r)$ are illustrated in Figs.~1 and~2.
It is observed that the functions $f=f(r)$, $h=h(r)$, $h'=h'(r)$, $P=P(r)$, and $r_*=r_*(r)$ are all continuous at the surface of the star $r=r_s$.
By assuming uniform density, $\rho$ is constant inside the star. 
In the next section, we will explore the axial gravitational perturbations of the obtained metric.

\begin{figure*}[htbp]
\centering
\includegraphics[width=0.8\columnwidth]{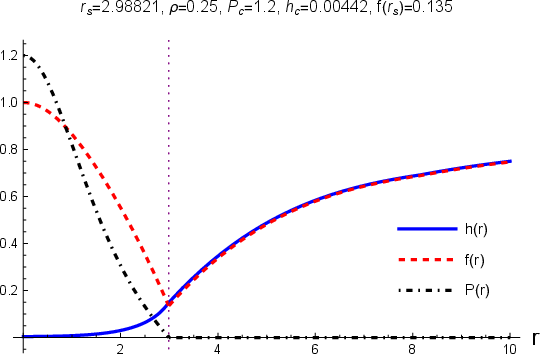}\includegraphics[width=0.8\columnwidth]{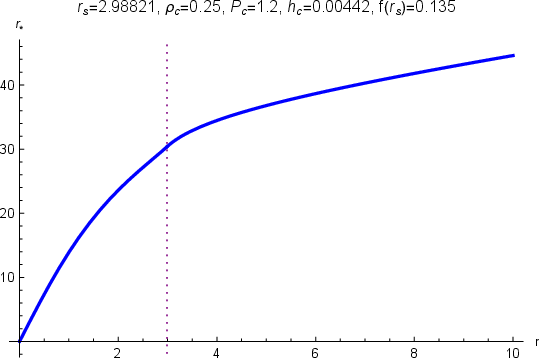}
\caption{Left: Profiles of $f=f(r)$, $h=h(r)$, and $P=P(r)$ as functions of the radial coordinate.
Right: The relation between the radial and tortoise coordinates.
The calculations are performed using $c_{14}=0.1$, $\rho=0.25$, and $P_c=1.2$.
}
\lb{Fig1}
\end{figure*}

\begin{figure*}[htbp]
\centering
\includegraphics[width=0.8\columnwidth]{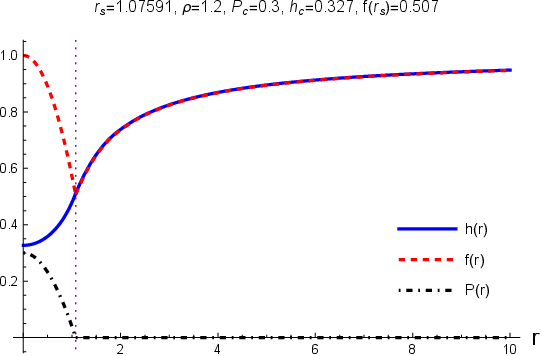}\includegraphics[width=0.8\columnwidth]{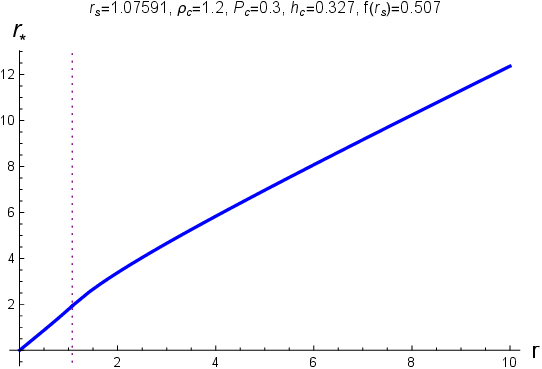}
\caption{The same as Fig.~\ref{Fig1} but for the parameters $c_{14}=0.1$, $\rho=1.2$, and $P_c=0.3$.
}
\lb{Fig2}
\end{figure*}

\section{Axial gravitational perturbation and Numerical Results}
\renewcommand{\theequation}{3.\arabic{equation}} \setcounter{equation}{0}

To investigate the axial gravitational perturbations, we consider the Regge-Wheeler gauge
\bqn
\lb{metric3}
\delta g_{\mu\nu}&=&
\begin{bmatrix}
0 & 0 & 0 & h_0(r) \\
0 & 0 & 0 & h_1(r) \\
0 & 0 & 0 & 0  \\
h_0(r) & h_1(r) & 0 & 0
\end{bmatrix}
e^{-i\omega t}\sin\theta\partial_\theta P_L(\cos\theta)\nb\\
\delta u_\mu&=&\delta^\varphi_\mu h_n(r)e^{-i\omega t}\sin\theta\partial_\theta P_L(\cos\theta)
\eqn
where the method of separation of variables is adopted, it suffices to consider the vanishing magnetic quantum number $M=0$.
Also, we consider $L=2$ in our numerical calculations.
For perturbations of axial parity, Regge and Wheeler use the freedom to choose $h_1$ as the gauge-invariant quantity~\cite{Nollert:1999ji}.
Subsequently, one results in two independent degrees of freedom $h_1$ and $h_n$.
The backreaction is ignored as the perturbations are insignificant, namely $g_{\mu\nu}\gg\delta g_{\mu\nu}$ and $u_{\mu}\gg\delta u_{\mu}$.

By introducing the transformation
\bqn
\lb{transformation}
h_1(r)&=&r^2e^{\int\left[\frac{(1-2c_{14})rP-r\rho}{(c_{14}-2)f}-\frac{1}{rf}\right] dr}R_B(r) ,\nb\\
h_n(r)&=&R_C(r)/\omega ,
\eqn
we find two coupled master equations for axial gravitational perturbations as follows
\bqn
\lb{masterequation1}
&&\sqrt{f(r)h(r)}\frac{d}{dr}\left(\sqrt{f(r)h(r)}\frac{d}{dr}R_B(r)\right)\nb\\
&&~~~~~+\left(\frac{\omega^2}{c_T^2}-V_T(r)\right)R_B(r)=U_T(r)R_C(r)\nb\\
&&\sqrt{f(r)h(r)}\frac{d}{dr}\left(\sqrt{f(r)h(r)}\frac{d}{dr}R_C(r)\right)\nb\\
&&~~~~~+\left(\frac{\omega^2}{c_V^2}-V_V(r)\right)R_C(r)=U_V(r)R_B(r)
\eqn
where the effective potentials $V_T$, $V_V$, $U_T$, and $U_V$ are given in Appx.~\ref{appB}, while the propagation speeds of the vector and tensor sectors, $c_V$ and $c_T$, are given by 
\bqn
\lb{cij}
c_V^2&=&\frac{2c_1-c_{+}(2c_1-c_{+})}{2c_{14}(1-c_{+})},\nb\\
c_T^2&=&\frac{1}{1-c_{+}}.
\eqn
For convenience, the tortoise coordinate is introduced as $r_*=\int_0^rdr/\sqrt{fh}$.
One notes that the derivation of Eq.~(\ref{masterequation1}) is based on the feasibility of separating variables.

As the effective potentials vanish at infinity and diverge at the center, the boundary conditions read
\bqn
\lb{Boundary}
R_B(r)&\sim&
\left\{
  \begin{array}{cc}
    r^{1+L} & r\rightarrow 0\\
    e^{i\frac{\omega}{c_T} r_*}  & r\rightarrow\infty \\
  \end{array}
\right.\nb\\
R_C(r)&\sim&
\left\{
  \begin{array}{cc}
    r^{1+L} & r\rightarrow 0\\
    e^{i\frac{\omega}{c_V} r_*}  & r\rightarrow\infty \\
  \end{array}
\right.
\eqn

In order to study the spacetime evolution of the coupled axial gravitational perturbations, one can work out the master equation in a similar fashion in the time domain and employ the finite difference method~\cite{FDM1, FDM2, FDM3, FDM4, FDM5, FDM6, FDM7}.
Specifically, we set $t_i=t_0+i\Delta t$, $r_{*j}=j\Delta r_*$, $\psi^i_j=R_B(t=t_i,r_*=r_{*j})$, $\phi^i_j=R_C(t=t_i,r_*=r_{*j})$, $V_j^T=V_T(r_*=r_{*j})$, $V_j^V=V_V(r_*=r_{*j})$, $U_j^T=U_T(r_*=r_{*j})$ and $U_j^V=U_V(r_*=r_{*j})$. 
The discretized master equation reads
\bqn
\lb{Numerical3}
\psi^{i+1}_j&=&-\psi^{i-1}_j+\frac{\Delta t^2}{c_T^2\Delta r_*^2}\left(\psi^i_{j-1}+\psi^i_{j+1}\right)\nb\\
&&+\left(2-2\frac{\Delta t^2}{c_T^2\Delta r_*^2}-\frac{\Delta t^2}{c_T^2} V^T_j\right)\psi^i_j-\frac{\Delta t^2}{c_T^2} U^T_j\phi^i_j ,\nb\\
\phi^{i+1}_j&=&-\phi^{i-1}_j+\frac{\Delta t^2}{c_V^2\Delta r_*^2}\left(\phi^i_{j-1}+\phi^i_{j+1}\right)\nb\\
&&+\left(2-2\frac{\Delta t^2}{c_V^2\Delta r_*^2}-\frac{\Delta t^2}{c_V^2} V^V_j\right)\phi^i_j-\frac{\Delta t^2}{c_V^2} U^V_j\psi^i_j ,\nb\\
\eqn
for which the boundary conditions are:
\bqn
\lb{Numerical4}
\psi^i_0=\phi^i_0=0.
\eqn
On the star's surface $r_{*s}\equiv r_*(r_s)=N_s\Delta r_*$, the connection condition is discretized into the following form
\bqn
\lb{Numerical6}
\psi^i_{N_s}=\left(\psi^i_{N_s-1}+\psi^i_{N_s+1}\right)/2,\nb\\
\phi^i_{N_s}=\left(\phi^i_{N_s-1}+\phi^i_{N_s+1}\right)/2.
\eqn
The spacetime evolution of the axial gravitational perturbations can be carried out using the finite difference method can be performed by specifying the initial conditions for $R_B^0(x_j)$, $\dot R_B^0(x_j)$, $R_C^0(x_j)$ and $\dot R_C^0(x_j)$:
\bqn
\lb{Numerical5}
\psi^0_j&=&R_B^0(r_*=r_{*j}),\nb\\
\psi^1_j&=&\psi^0_j+\Delta t \dot R_B^0(r_*=r_{*j}),\nb\\
\phi^0_j&=&R_C^0(r_*=r_{*j}),\nb\\
\phi^1_j&=&\psi^0_j+\Delta t \dot R_C^0(r_*=r_{*j}).
\eqn

\begin{figure*}[htbp]
\centering
\includegraphics[width=0.8\columnwidth]{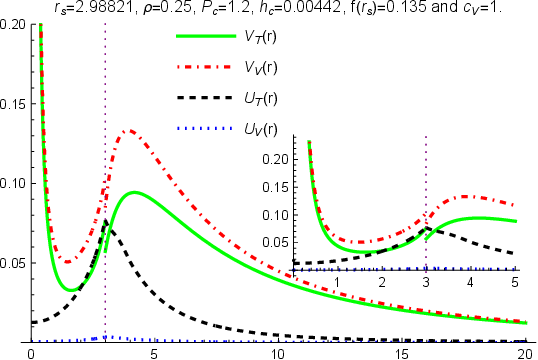}\includegraphics[width=0.8\columnwidth]{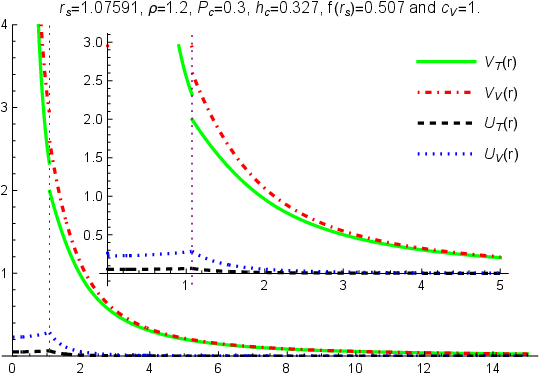}
\includegraphics[width=0.8\columnwidth]{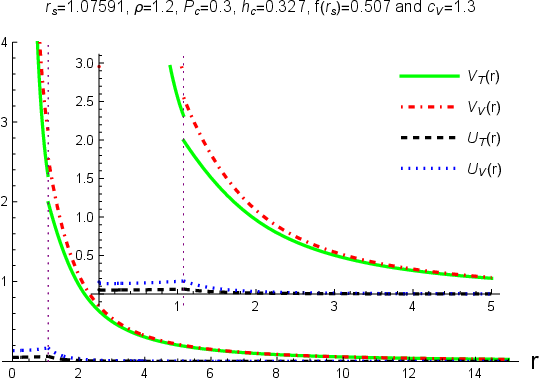}\includegraphics[width=0.8\columnwidth]{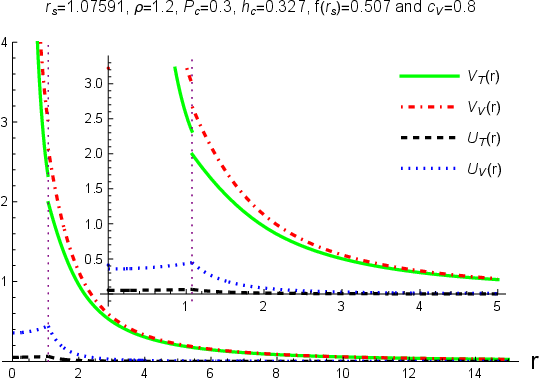}
\caption{The effective potentials $V_T=V_T(r)$, $U_T=U_T(r)$, $V_V=V_V(r)$, and $U_V=U_V(r)$ evaluated using different parameters.
Top-left: The resulting effective potential with a potential well, obtained by assuming $c_T=c_V=1$, $c_{14}=0.1$, $\rho=0.25$, and $P_c=1.2$. 
Top-right: The resulting effective potential, which decreases exponentially with the radial coordinate, obtained by assuming $c_T=c_V=1$, $c_{14}=0.1$, $\rho=1.2$, and $P_c=0.3$, where the two propagation speeds are identical.
Bottom-left: Similar to the top-right panel but obtained using the parameters $c_T=1$, $c_V=1.3$, $c_{14}=0.1$, $\rho=1.2$, and $P_c=0.3$, where $c_T > c_V$.
Bottom-right: Similar to the top-right panel but obtained using the parameters $c_T=1$, $c_V=0.8$, $c_{14}=0.1$, $\rho=1.2$, and $P_c=0.3$, where $c_T < c_V$. }
\lb{Fig3}
\end{figure*}

In Fig.~3, we represent the effective potentials for different scenarios.
In the top-left panel, it is observed that the effective potential features a potential well for both $V_T$ and $V_V$ by assuming the parameters so that the radius of the star is smaller than the maximum of the vacuum Schwarzschild-type solution.
Conversely, as the star's radius increases and becomes more significant than the {\it could-have-been} maximum, the effective potential decreases monotonically with the radial coordinate.
We consider three cases for the latter scenario based on the relative sizes of the propagation speeds of the bimodal medium, as shown in the top-right, bottom-left, and bottom-right panels.
A discontinuity is present at the star's surface in all these cases.
However, as elaborated below, while the discontinuity does not have a sizable impact on the waveforms for the scenario where the effective potential possesses a potential well, it plays a crucial role in the cases where the effective potential does not have a local maximum.

\begin{figure*}[htbp]
\centering
\includegraphics[width=0.8\columnwidth]{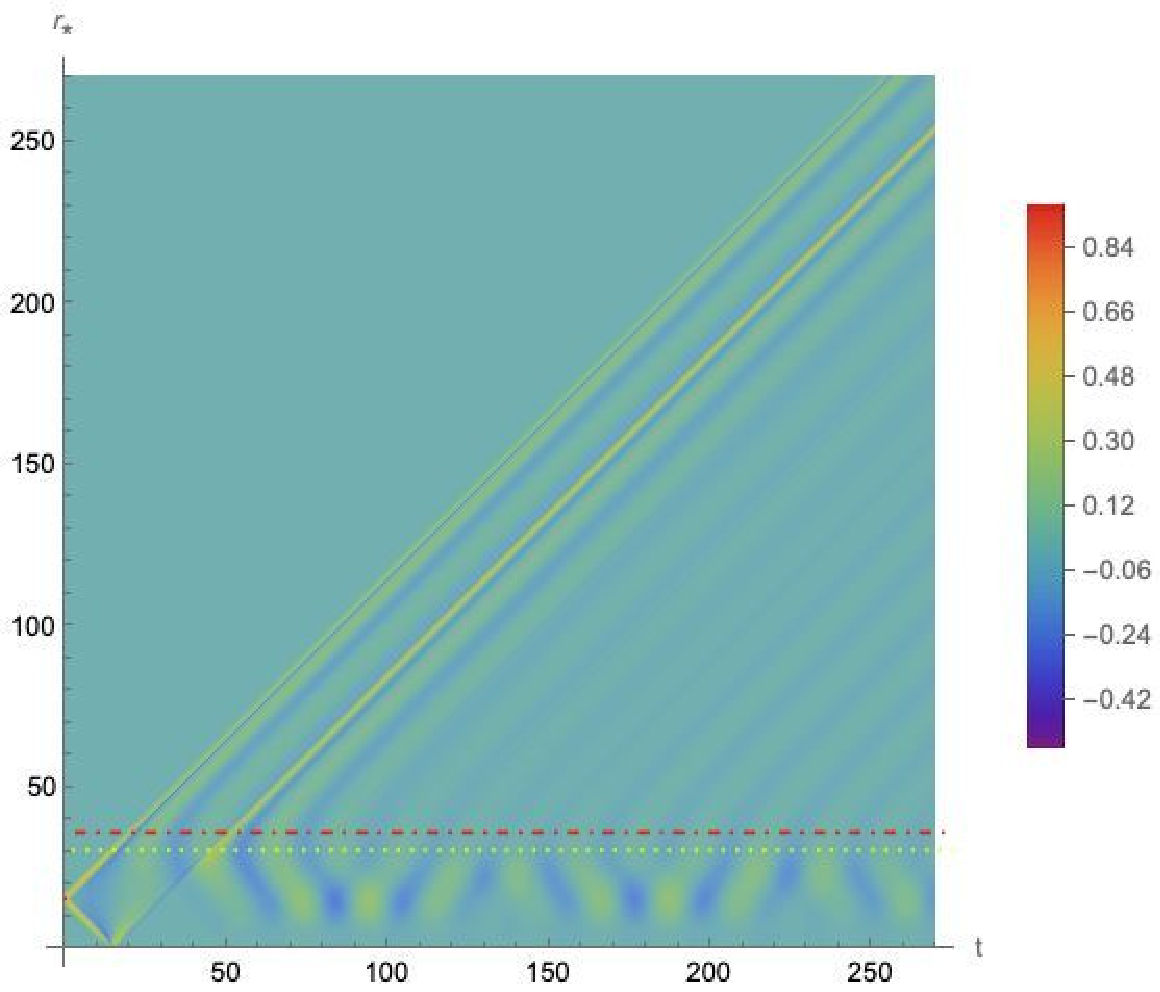}\includegraphics[width=0.8\columnwidth]{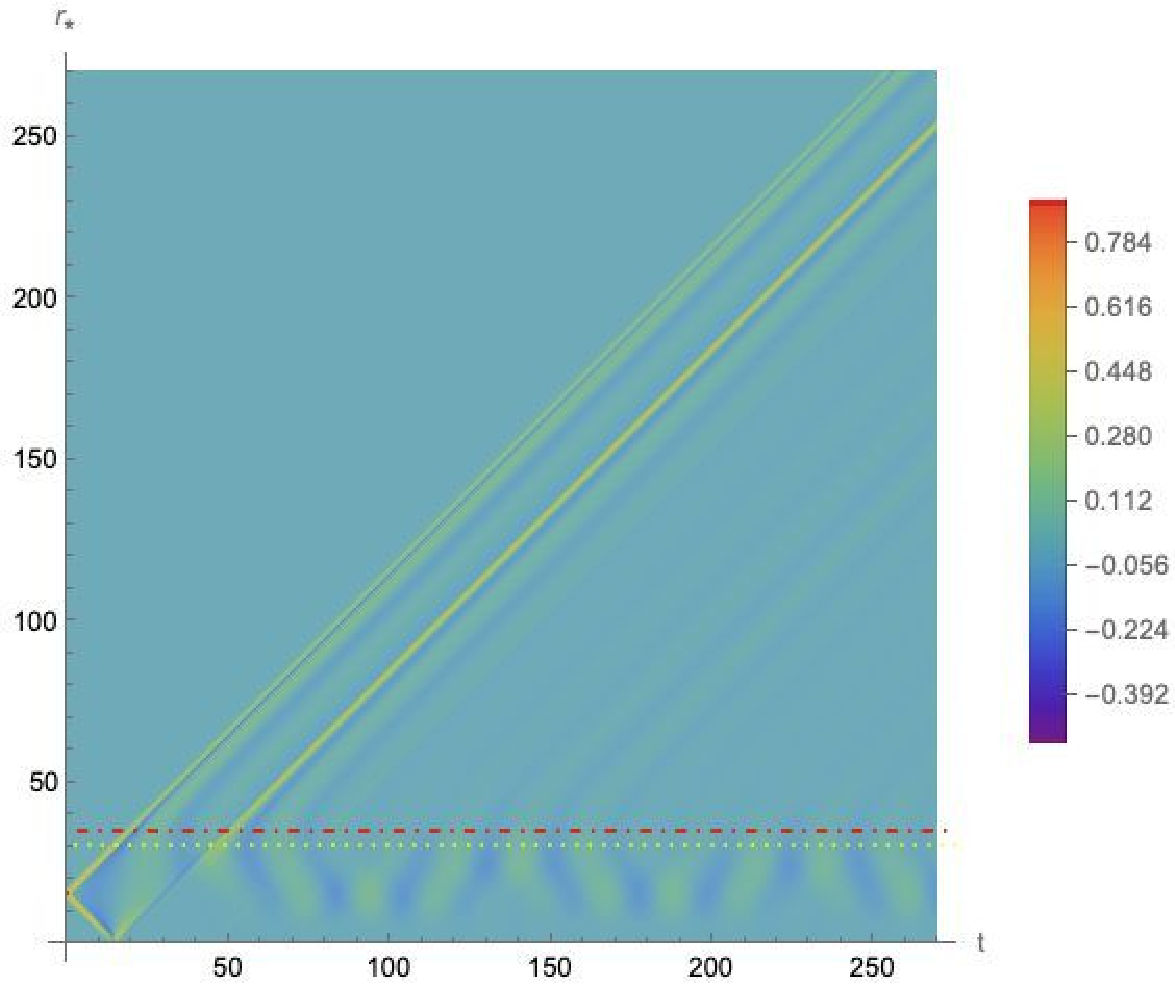}
\includegraphics[width=0.8\columnwidth]{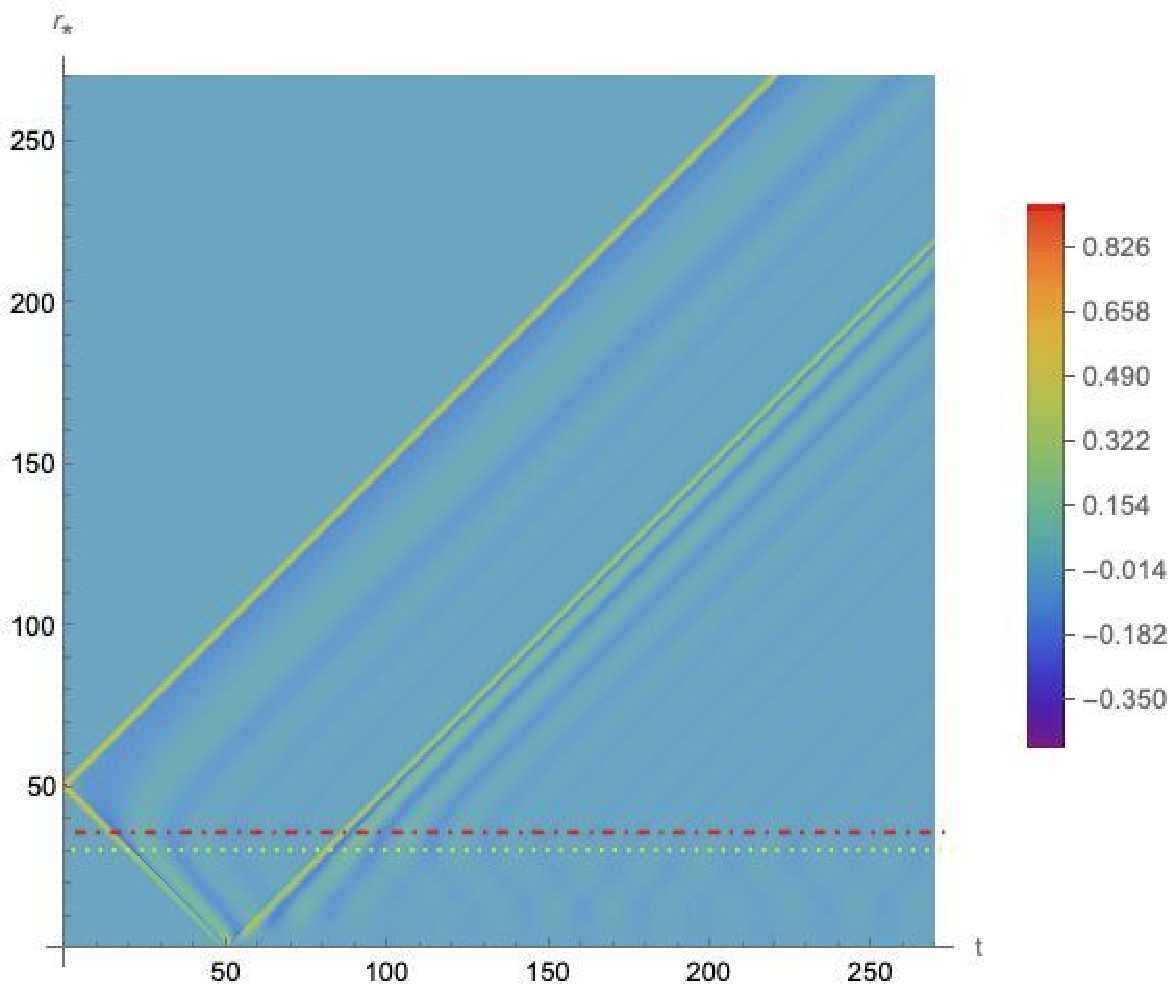}\includegraphics[width=0.8\columnwidth]{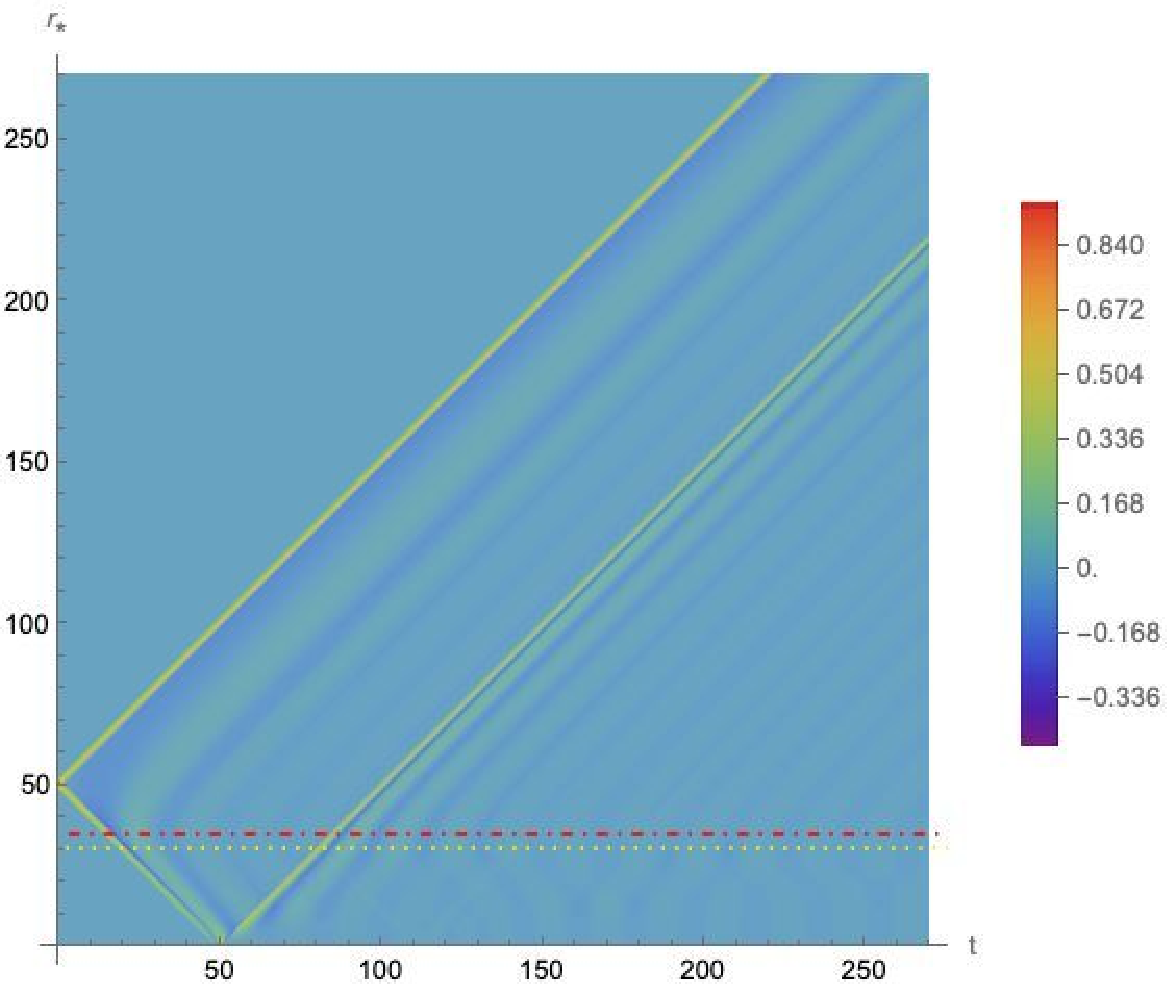}
\caption{Numerical results of the spacetime evolutions of $R_B$ (left column) and $R_C$ (right column).
The two rows represent results obtained using different initial perturbations, as discussed in the text.
The calculations are carried out using the parameters $c_T=c_V=1$, $c_{14}=0.1$, $\rho=0.25$, and $P_c=1.2$, corresponding to the effective potentials shown in the top-left panel of Fig.~3.
The yellow dotted horizontal lines correspond to the star's surface, while the red dashed ones indicate the maximum of the effective potential.
}
\lb{Fig4}
\end{figure*}

\begin{figure*}[htbp]
\centering
\includegraphics[width=0.8\columnwidth]{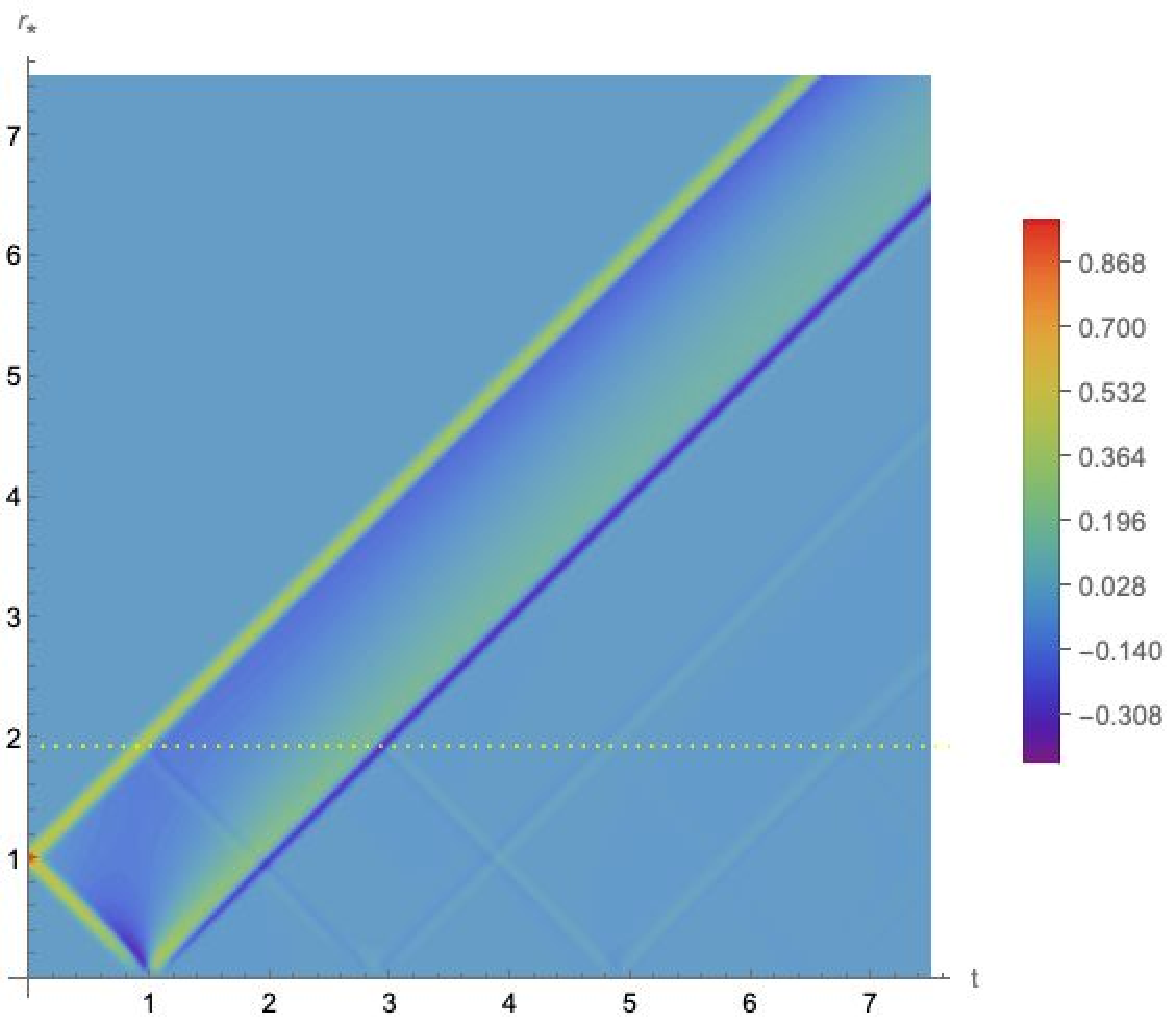}\includegraphics[width=0.8\columnwidth]{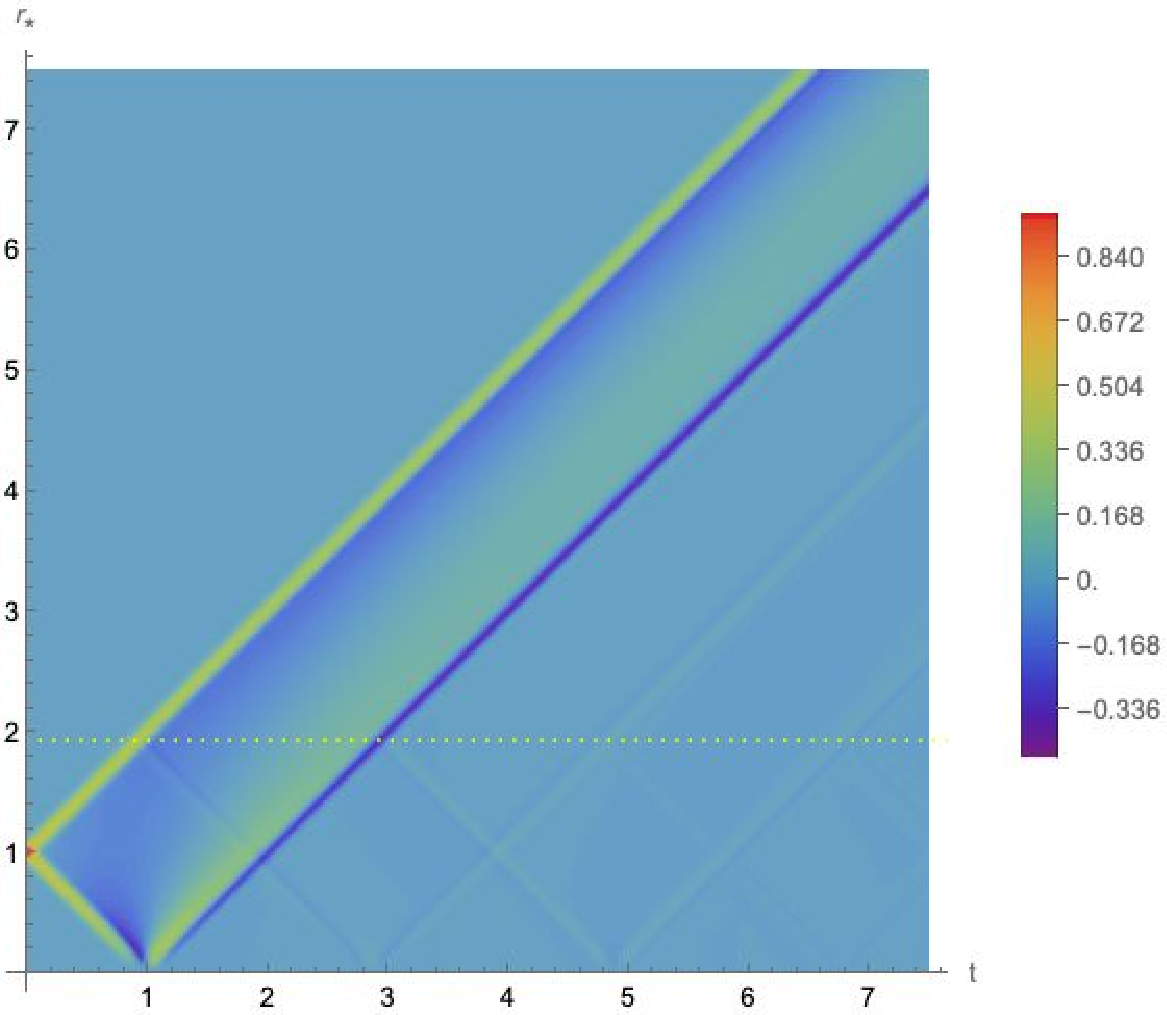}
\includegraphics[width=0.8\columnwidth]{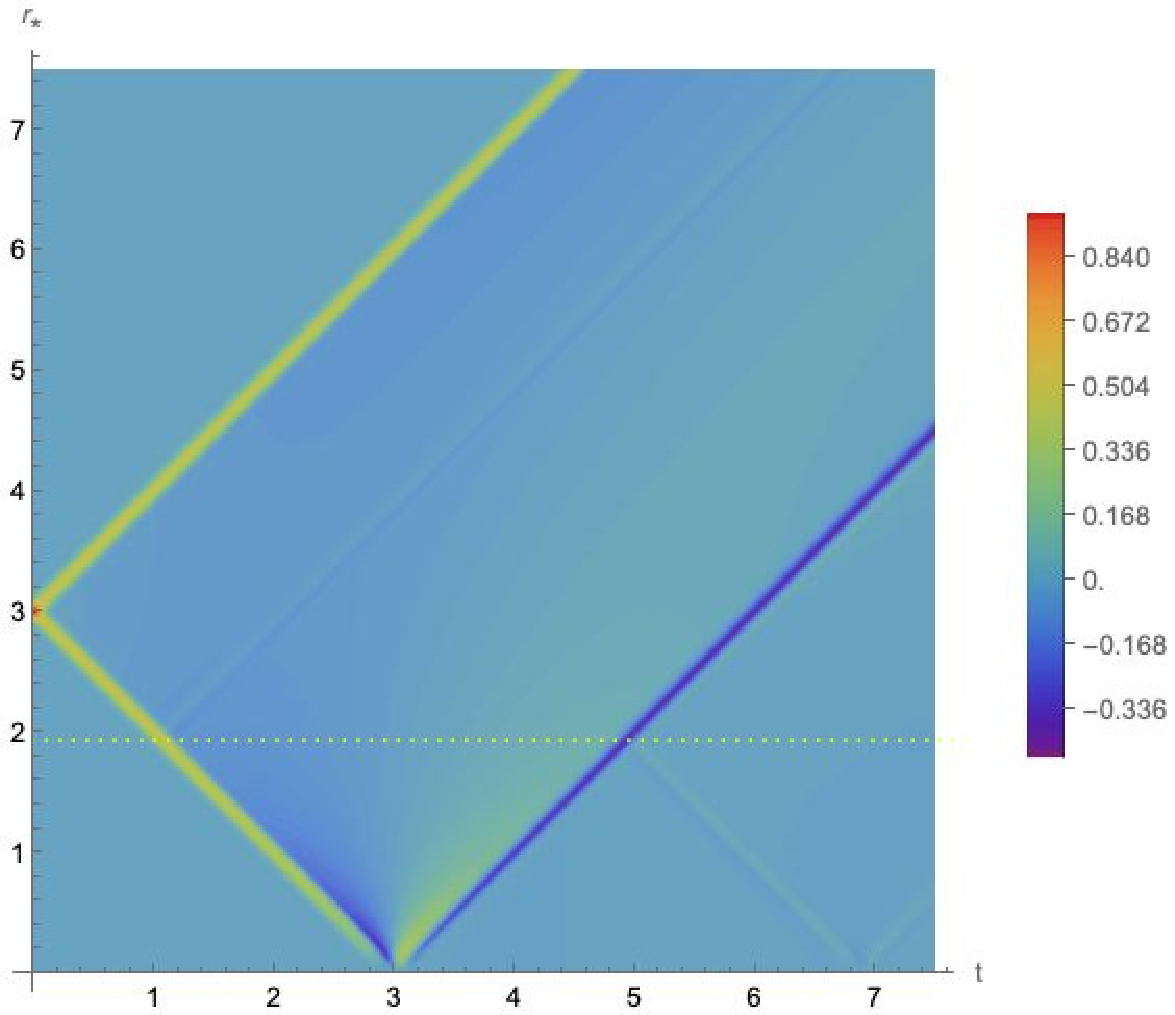}\includegraphics[width=0.8\columnwidth]{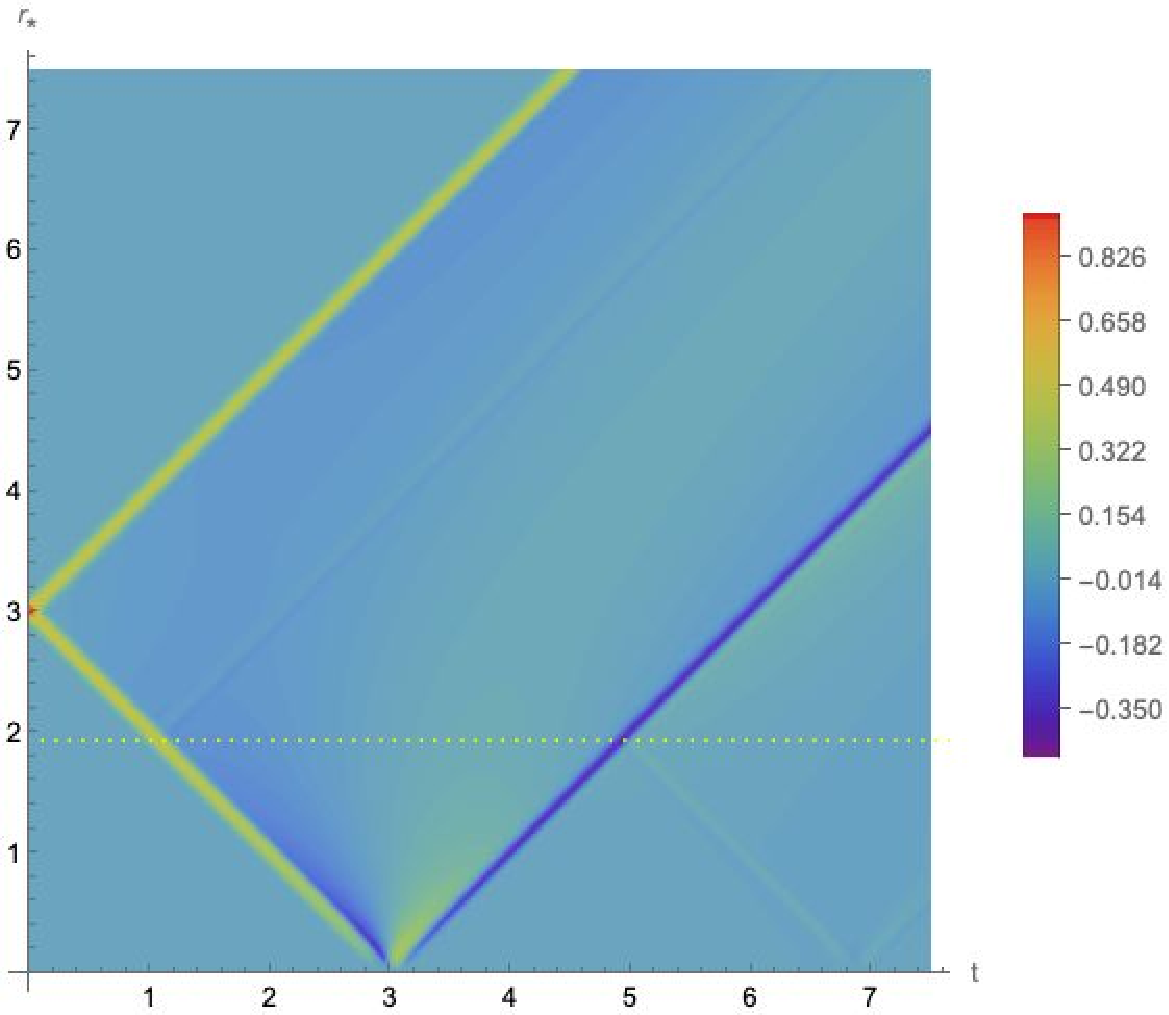}
\caption{The same as Fig.~4, but for the parameters $c_T=c_V=1$, $c_{14}=0.1$, $\rho=1.2$, and $P_c=0.3$, associated with the effective potentials shown in the top-right panel of Fig.~3.
}
\lb{Fig5}
\end{figure*}

\begin{figure*}[htbp]
\centering
\includegraphics[width=0.8\columnwidth]{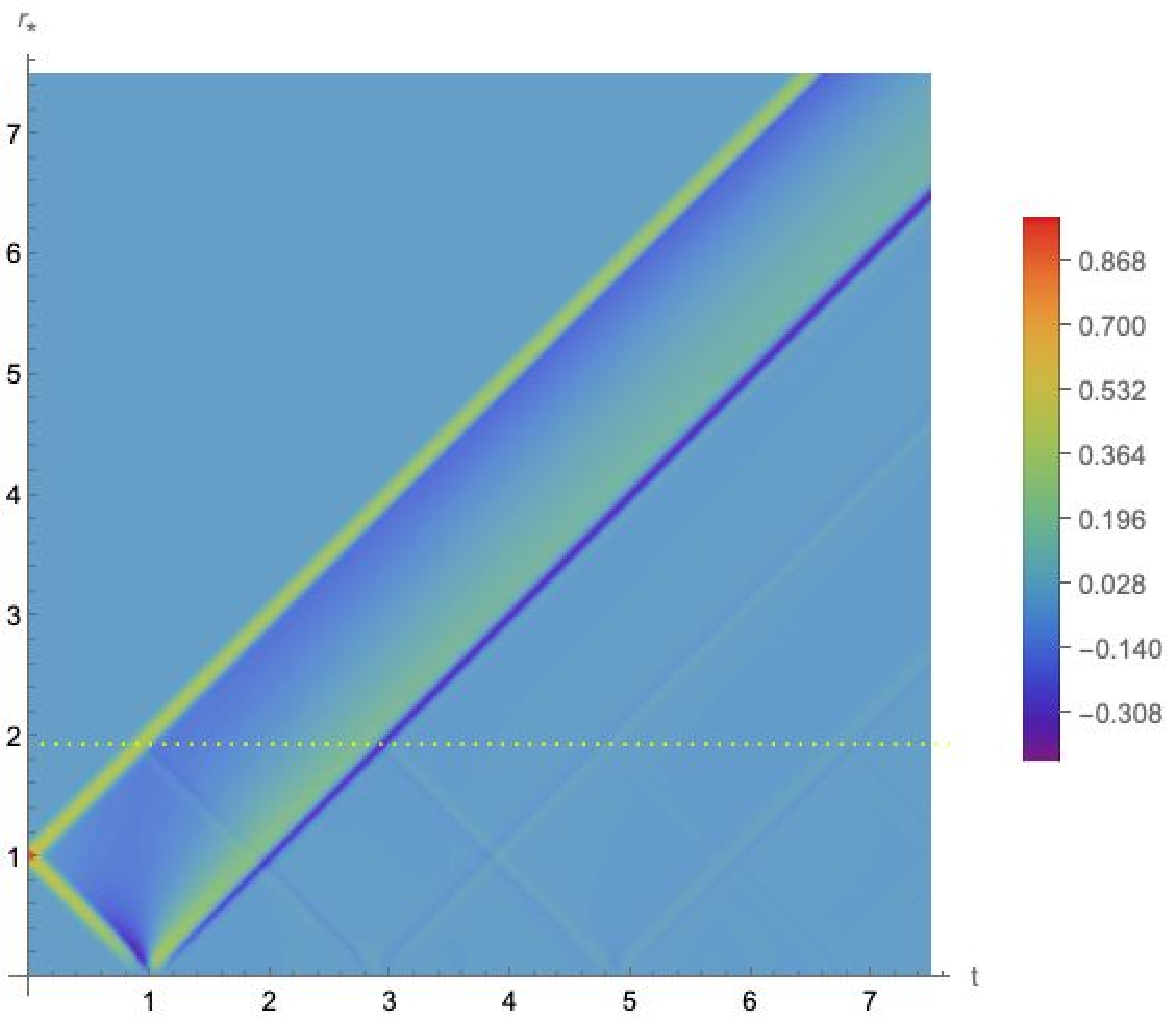}\includegraphics[width=0.8\columnwidth]{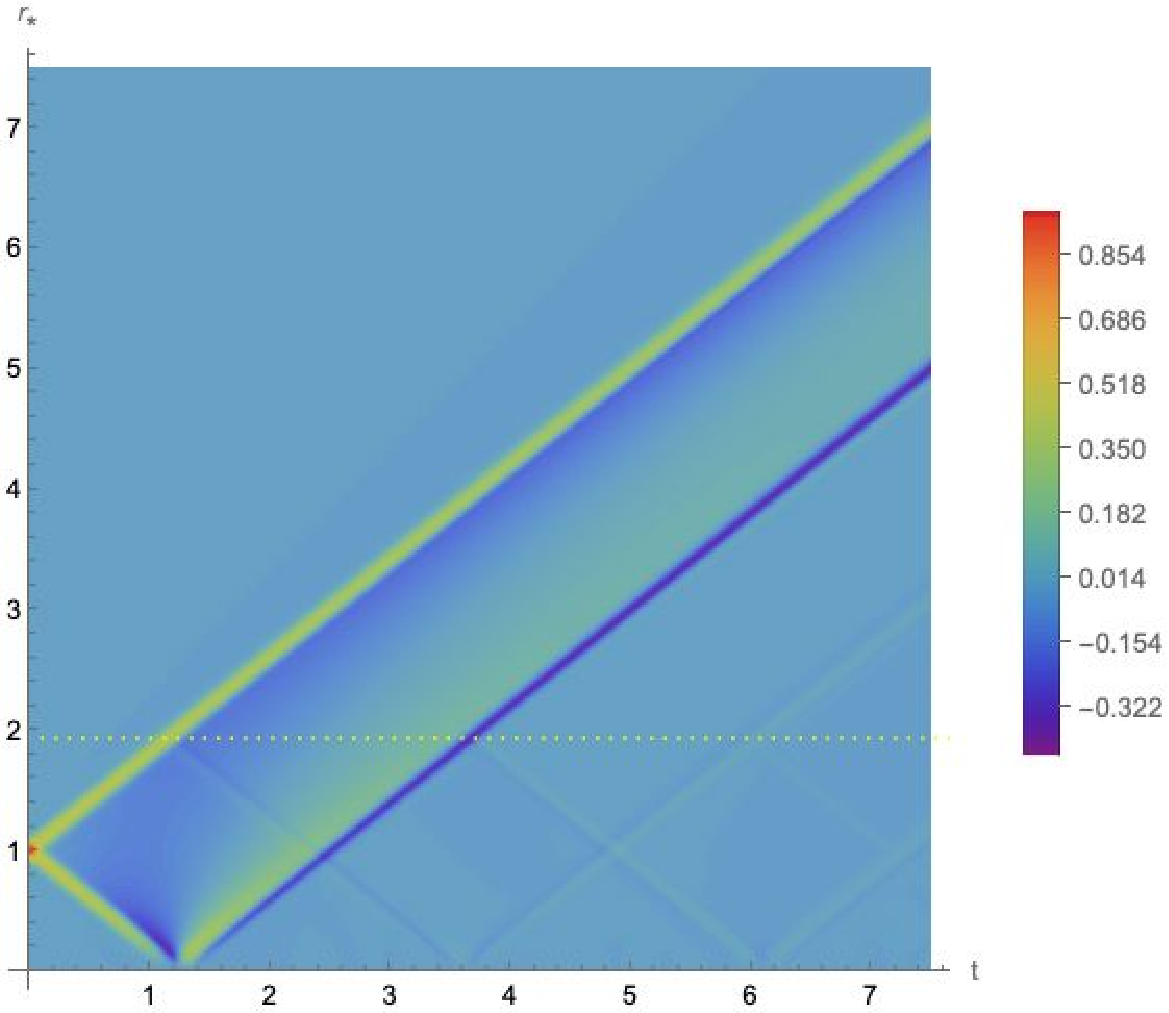}
\includegraphics[width=0.8\columnwidth]{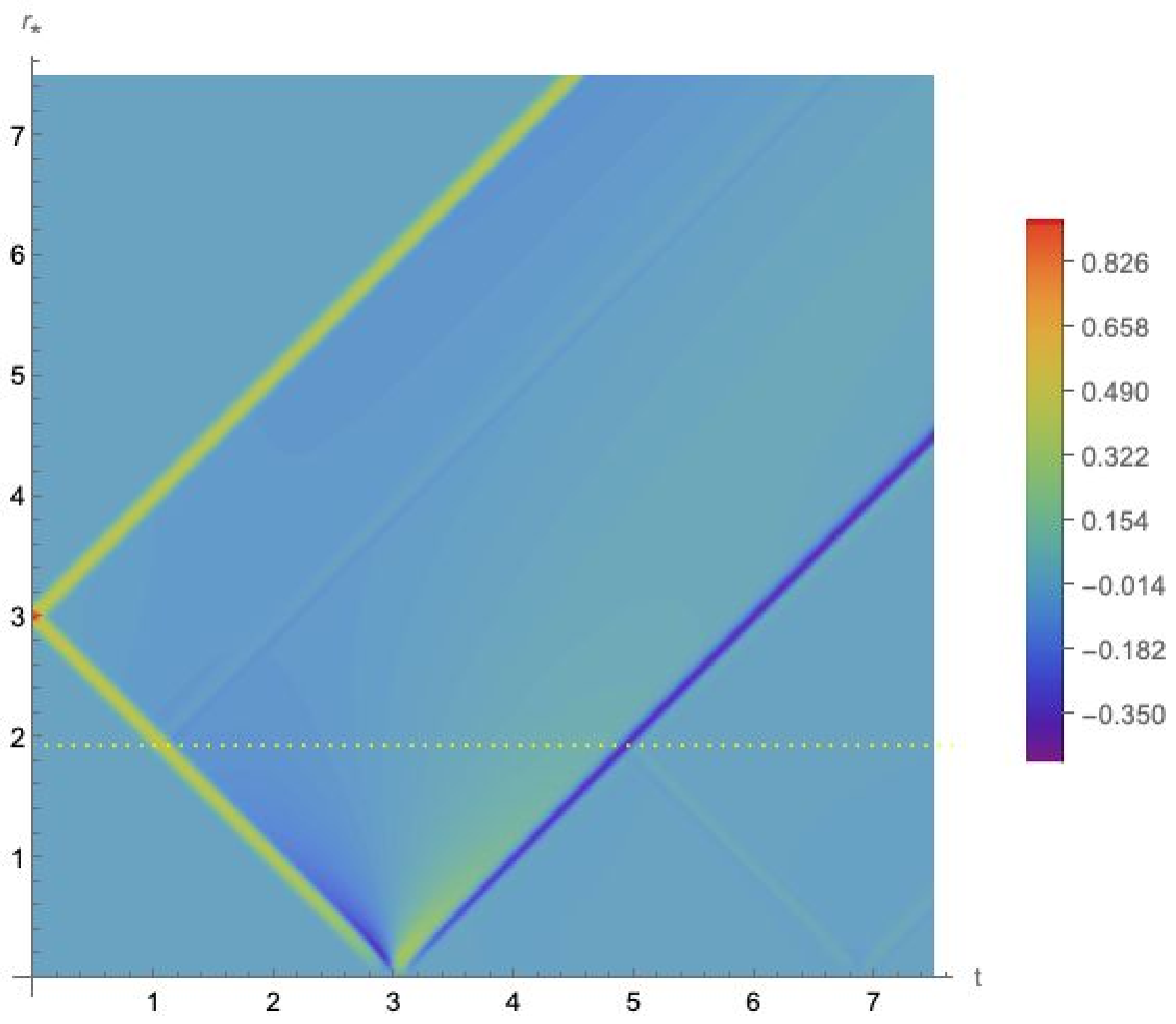}\includegraphics[width=0.8\columnwidth]{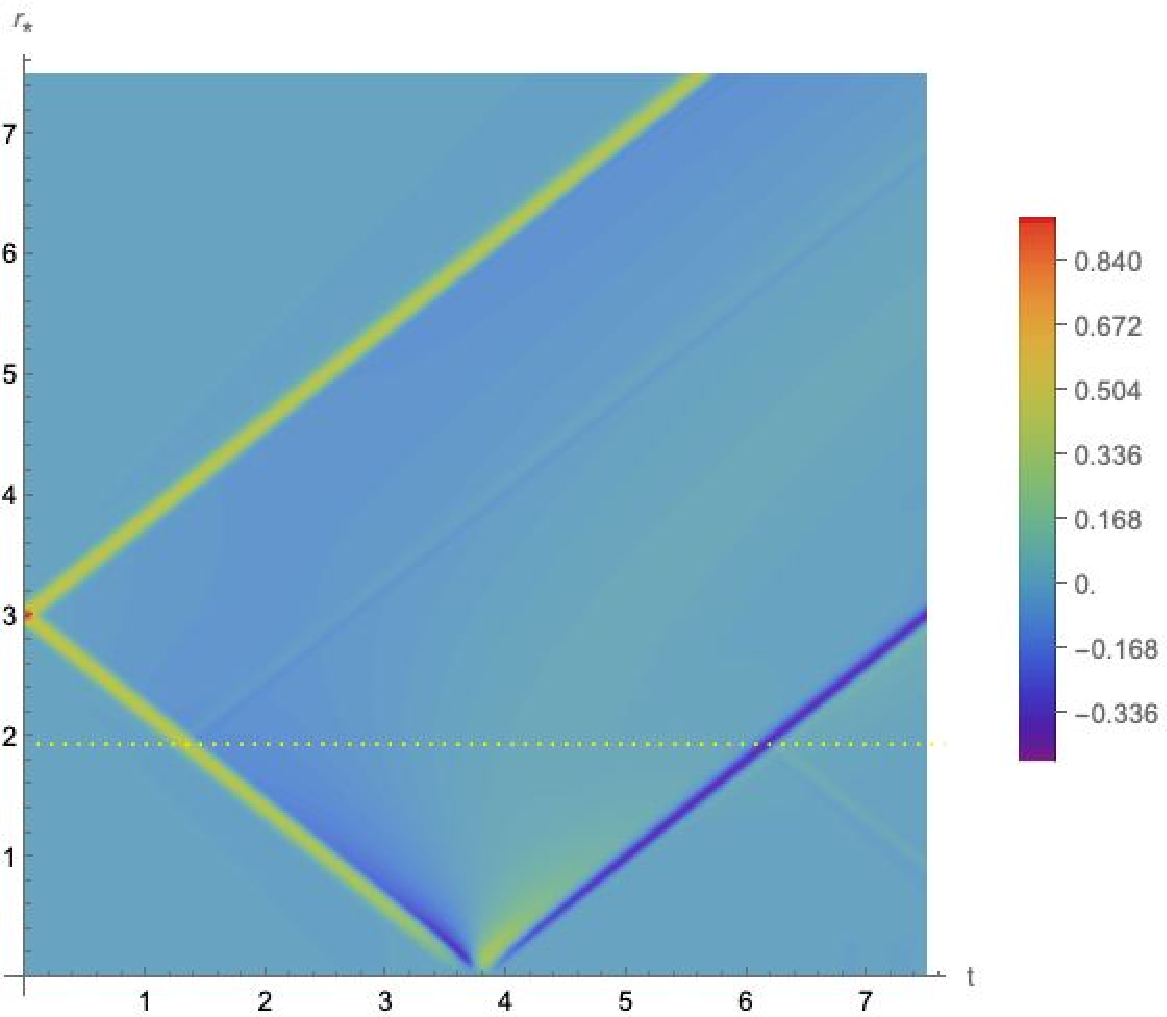}
\caption{The same as Fig.~4, but for the parameters $c_T=1$, $c_V=0.8$, $c_{14}=0.1$, $\rho=1.2$, and $P_c=0.3$, associated with the effective potentials shown in the bottom-left panel of Fig.~3.
}
\lb{Fig6}
\end{figure*}

\begin{figure*}[htbp]
\centering
\includegraphics[width=0.8\columnwidth]{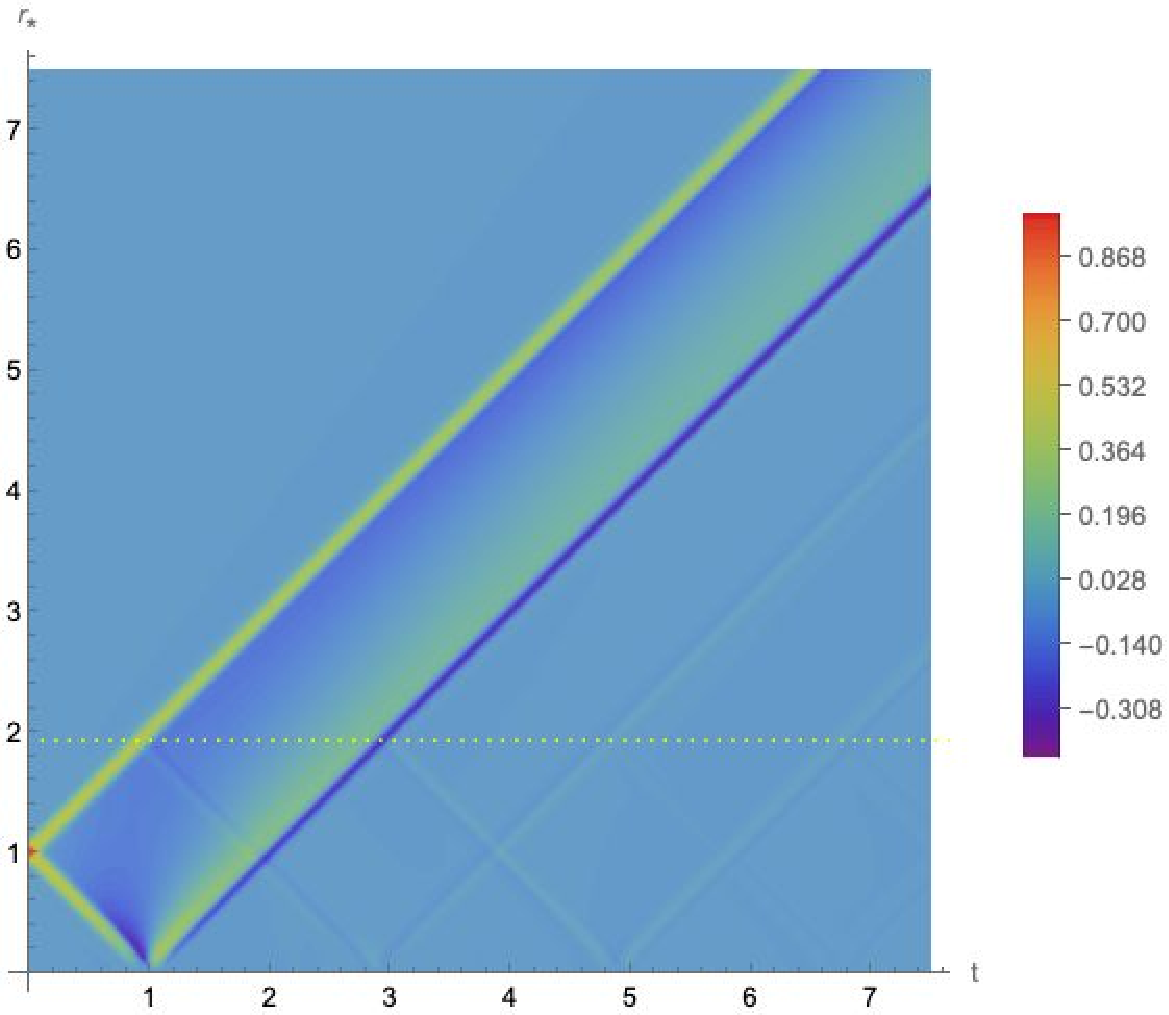}\includegraphics[width=0.8\columnwidth]{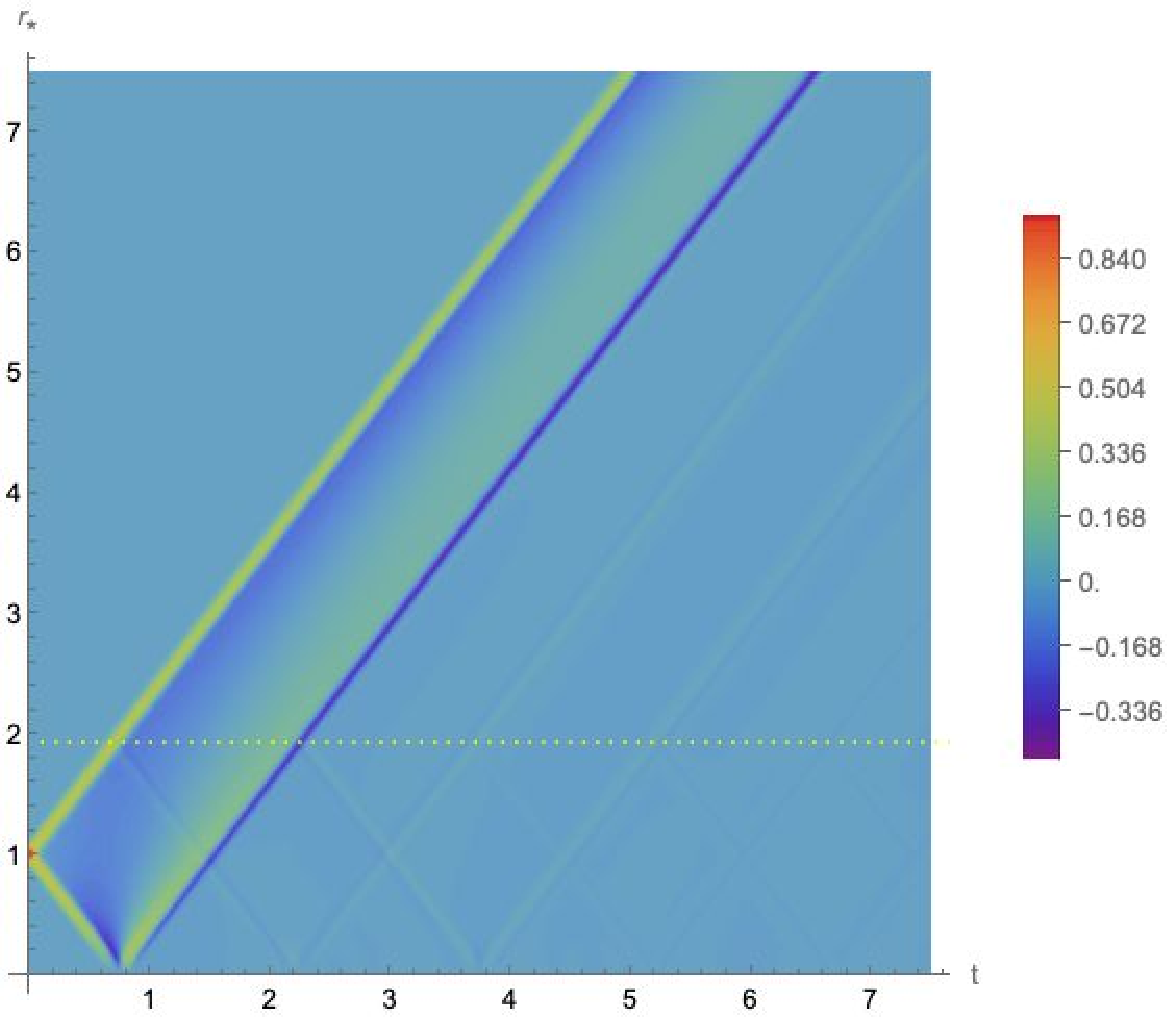}
\includegraphics[width=0.8\columnwidth]{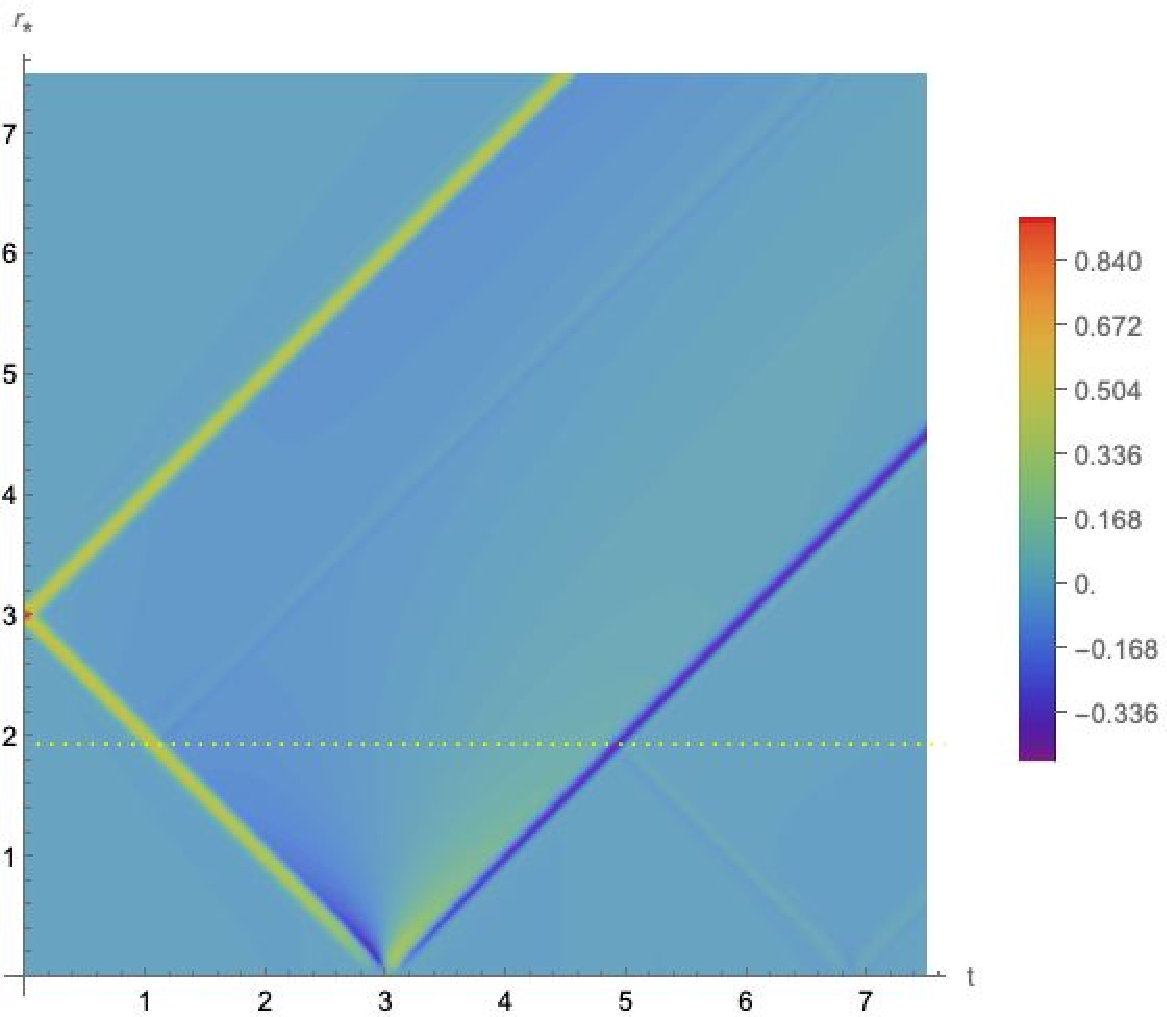}\includegraphics[width=0.8\columnwidth]{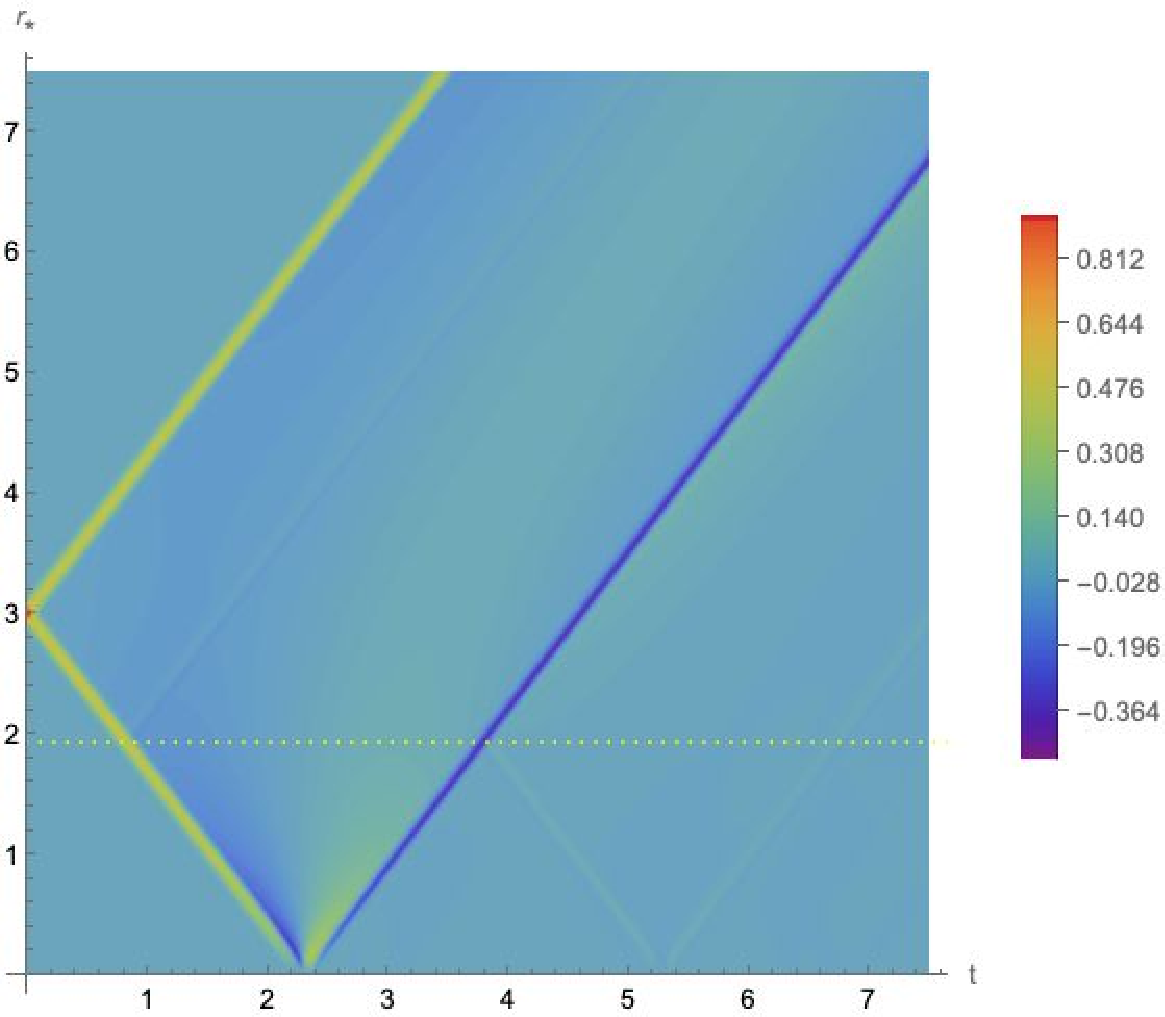}
\caption{The same as Fig.~4, but for the parameters $c_T=1$, $c_V=1.3$, $c_{14}=0.1$, $\rho=1.2$, and $P_c=0.3$, associated with the effective potentials shown in the bottom-right panel of Fig.~3.
}
\lb{Fig7}
\end{figure*}

In Fig.~4, we show the resulting spacetime evolutions of the axial gravitational perturbations.
Without loss of generality, in our calculations, we consider stars with a given radius $r_*(r_s)=31.93$. 
For the initial conditions, we consider $R_B(t=0)=R_C(t=0)=e^{-2(r_*-r_{*C})^2}$ and $\dot{R}_B(t=0)=\dot{R}_C(t=0)=0$. 
For the upper row, we choose $r_{*C}=15 < r_*(r_s)$ so that the initial perturbations are placed inside the star. 
For the bottom row, we consider $r_{*C}=50 > r_*(r_s)$ so that the initial perturbations are located outside the star. 
Echoes emerge as the waves are repeatedly bounded back and forth between the potential's maximum (indicated by the dashed red horizontal line) and the star's center.
At the potential's maximum, both the transmission and reflection waves are observed, while the reflection part contributes to the echoes.
We note that such a phenomenon is consistently observed in all the panels of Fig.~4.
In other words, it persists for both degrees of freedom associated with the effective potential $V_T$ and $V_V$, as well as for both initial perturbations.
Specifically, if the initial perturbations reside outside the star, the transmission wave enters through the star and eventually triggers the echoes.
However, a visible reflection primarily takes place at the potential's maximum.
As discussed below, the discontinuity at the star's surface (indicated by the yellow dotted horizontal line) might also trigger echoes.
This explains why the reflection and transmission near the potential's maximum are less distinguishable than those shown in Fig.~5.
However, as the latter echoes are attenuated, the potential's maximum plays a more crucial role in forming the echoes.

In Fig.~5, we take $r_*(r_s)=1.94$ and adopt the initial conditions $R_B(t=0)=R_C(t=0)=e^{-200(r_*-r_{*C})^2}$ and $\dot{R}_B(t=0)=\dot{R}_C(t=0)=0$. 
For the upper row, we choose $r_{*C}=1 < r_*(r_s)$ so that the initial perturbations reside inside the star.  
For the bottom row, we consider $r_{*C}=3 > r_*(r_s)$ so that the initial perturbations are placed outside the star. 
As shown in the top-right panel of Fig.~3, the related effective potentials do not possess any local maximum near the star's surface.
As a result, the echoes are dictated by the star's center and surface.
This is indicated in all the panels of Fig.~5.
Again, this phenomenon persists for both degrees of freedom and is irrelevant to the placement of the initial perturbations.

For both scenarios, the echo period is governed by the ratio between the characteristic length and the propagation speed, namely, $T_\text{echo}=\text{relevant spatial distance}/\text{wave speed}$.

In Fig.~6 and~7, we choose $r_*(r_s)=1.94$ and the initial conditions $R_B(t=0)=R_C(t=0)=e^{-200(r_*-r_{*C})^2}$ and $\dot{R}_B(t=0)=\dot{R}_C(t=0)=0$. 
Again, we consider two different configurations for the initial condition, where one assigns $r_{*C}=1 < r_*(r_s)$ and $r_{*C}=3 > r_*(r_s)$, respectively, to the first and second rows of the figures.
For these two cases, the system is bimodal and possesses two different propagation speeds.
This is indicated by the difference in the slopes of the wave trajectories in the left and right columns of the figures.
Once again, echoes are observed, which can be attributed to the repeated bouncing of the perturbations between the center and surface of the star.
From Fig.~4 to~.7, one also observes a half-wave loss when the wave is reflected at the star's center. 
This occurs because the potential function at the center diverges, and the waveform must vanish as a boundary condition.
The latter is readily satisfied by considering a superposition with another wave moving toward a positive radial coordinate but with an opposite phase.
Such a phenomenon is an analogy when the oscillations propagate on a string whose other end is attached to a solid wall.
Owing to the distinct nature of wave propagation in a binodal system, it might be utilized to extract metric parameters closely associated with the observability of the propagation speed and echo period.

It is well-known that a shock wave might occur when a source moves faster than the medium's propagation speed.
For the present study, intuitively, a shock wave might be present in a bimodal medium.
Specifically, it seems plausible that the formation of a shock wave is triggered by the fast mode, acting as a source that moves at a superluminal speed. 
Nonetheless, such a scenario depends on the amplitude and initial conditions of the wave disturbances.
By assigning the initial perturbations exclusively to the fast mode, one might explore how the energy is transferred to the slow mode and whether a shock front can be formed.
However, while considering different model parameters and initial conditions, our numerical simulation does not support such speculation.
The initial waveform is found to be gradually suppressed in time without triggering any shock phenomenon.

Before closing this section, we discuss a particular scenario where the Einstein-{\AE}ther parameters attain the limit $c_i\to 0$.
Moreover, as pointed out in~\cite{Elliott:2005va}, the observation of ultra-high energy rays places a tight constraint on the model parameters.
Specifically, for the ultra-high energy cosmic rays to propagate for at least 10 Kpc before fragmentation owing to gravitational Cherenkov or virtual graviton-mediated processes, the model parameters must be extremely small.
In this case, one might speculate that the model essentially falls back to general relativity, leading to insignificant, if any, observational implications. 
In this regard, we argue that it is still plausible for the two modes governed by Eqs.~\eqref{cij} to be distinguished from their general relativity counterpart.
While the speed of the tensor mode always falls back to the unit at this limit, the vector mode $c_V$ might attain a different value at the limit $c_i\to 0$ by satisfying all the constraints established in~\cite{Elliott:2005va}.
Specifically, one may adopt some particular but otherwise arbitrary parameters that exhibit some features of the Einstein-{\AE}ther theory under such circumstances.
Numerical calculations have been carried out to demonstrate the above point, where one chooses the following set of parameters:
\bqn
&& c_1 = 1\times 10^{-40}, \nb\\
&& c_2 = \frac{9}{4}\times 10^{-40} - 1\times 10^{-60}\,\nb\\
&& c_3 = -2\times 10^{-40},\nb\\
&& c_4 = \frac14\times 10^{-40}, \label{parLimGR}
\eqn
which satisfies the above-mentioned requirements (c.f. Eqs.~(2.9 - 2.11) and (6.1 - 6.4) of~\cite{Elliott:2005va}).
One finds $c_T \to 1$ and $c_V^2\to \frac{c_1}{c_{14}} = 0.8$.
In Figs.~\ref{FigAppD2} and~\ref{FigAppD3}, we show the temporal and spacetime evolutions of the perturbations using the parameters given by Eq.~\eqref{parLimGR}.
The first column of Fig.~\ref{FigAppD2} shows the effective potentials associated with two different types of echo mechanisms, as discussed in Fig.~\ref{Fig3}.
For both cases, two distinct modes are presented in the middle and right columns, whose propagation velocities are governed by Eqs.~\eqref{cij}.
Besides, echoes associated with these setups are observed in the two rows of Fig.~\ref{FigAppD3}.
Therefore, one concludes that both echoes and two distinct modes may potentially persist in this limit.
However, it is important to note that the above feature associated with the vector degree of freedom can hardly lead to any observational effect.
The main reason is the smallness of the coupling~\cite{Elliott:2005va}, in terms of the coefficients $c_i$, between the vector field and the gravitational sector.
Since the detector is constituted by ordinary matter, it does not directly interact with the vector field; consequently, any nontrivial effect of the vector field (even if it exists, such as a distinct sound speed) will be significantly suppressed by these coefficients.
In fact, this can be formally demonstrated (as pointed out by the anonymous referee and summarized below in Appx.~C).
Specifically, as long as the matter field does not directly interact with the vector field and the vector field only weakly couples to the gravitational sector, the dynamics of the matter and gravity largely revert to their general relativity counterparts, even though the vector field may admit some nontrivial solutions.
In other words, the detector only senses the vector field via the gravitational field, which is further suppressed significantly by the smallness of the coefficients $c_i$, effectively eliminating any observable implications.

\begin{figure*}[htbp]
\centering
\includegraphics[width=0.7\columnwidth]{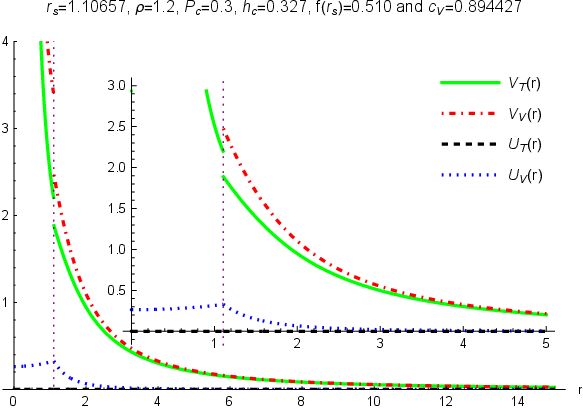}\includegraphics[width=0.7\columnwidth]{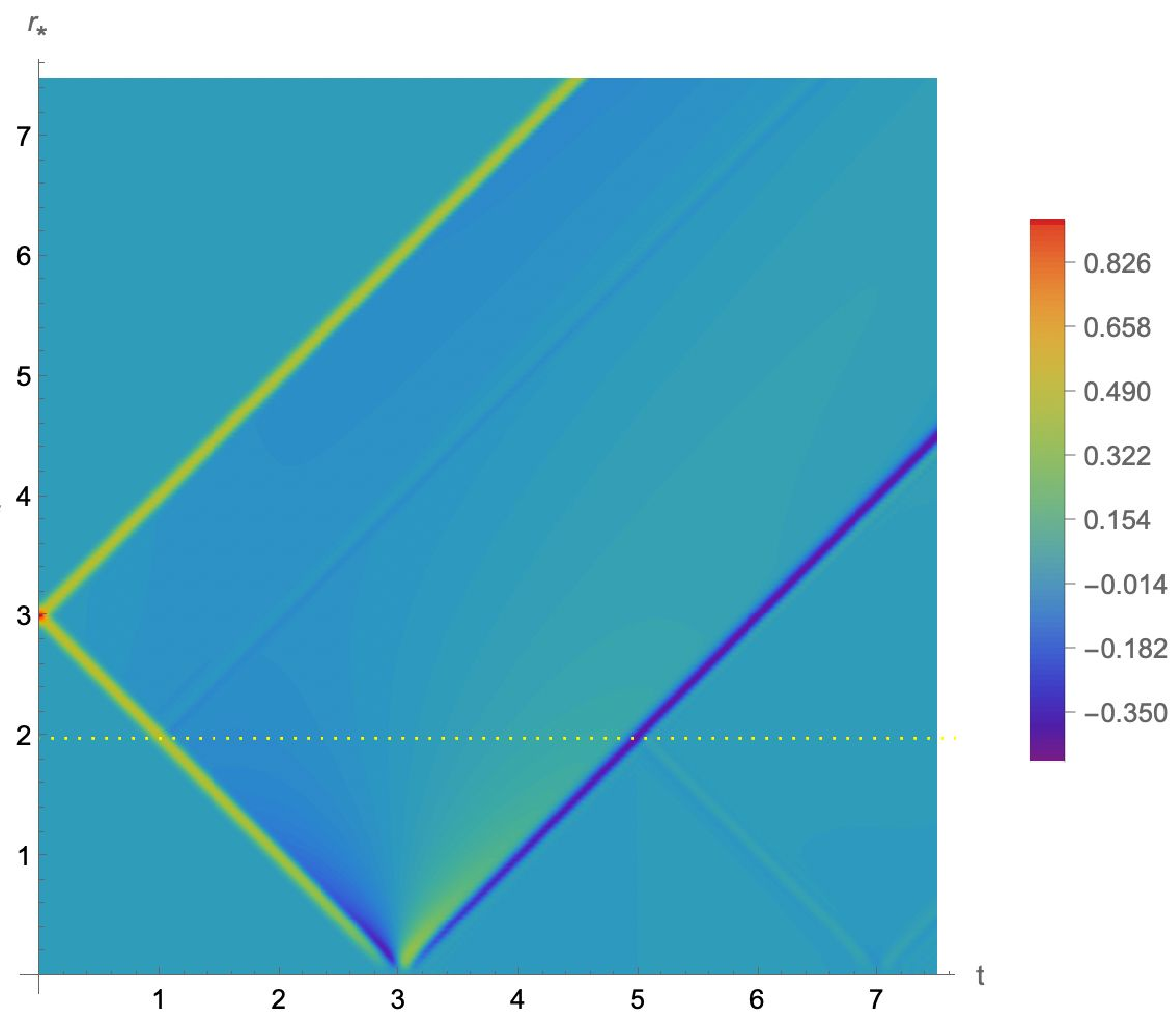}\includegraphics[width=0.7\columnwidth]{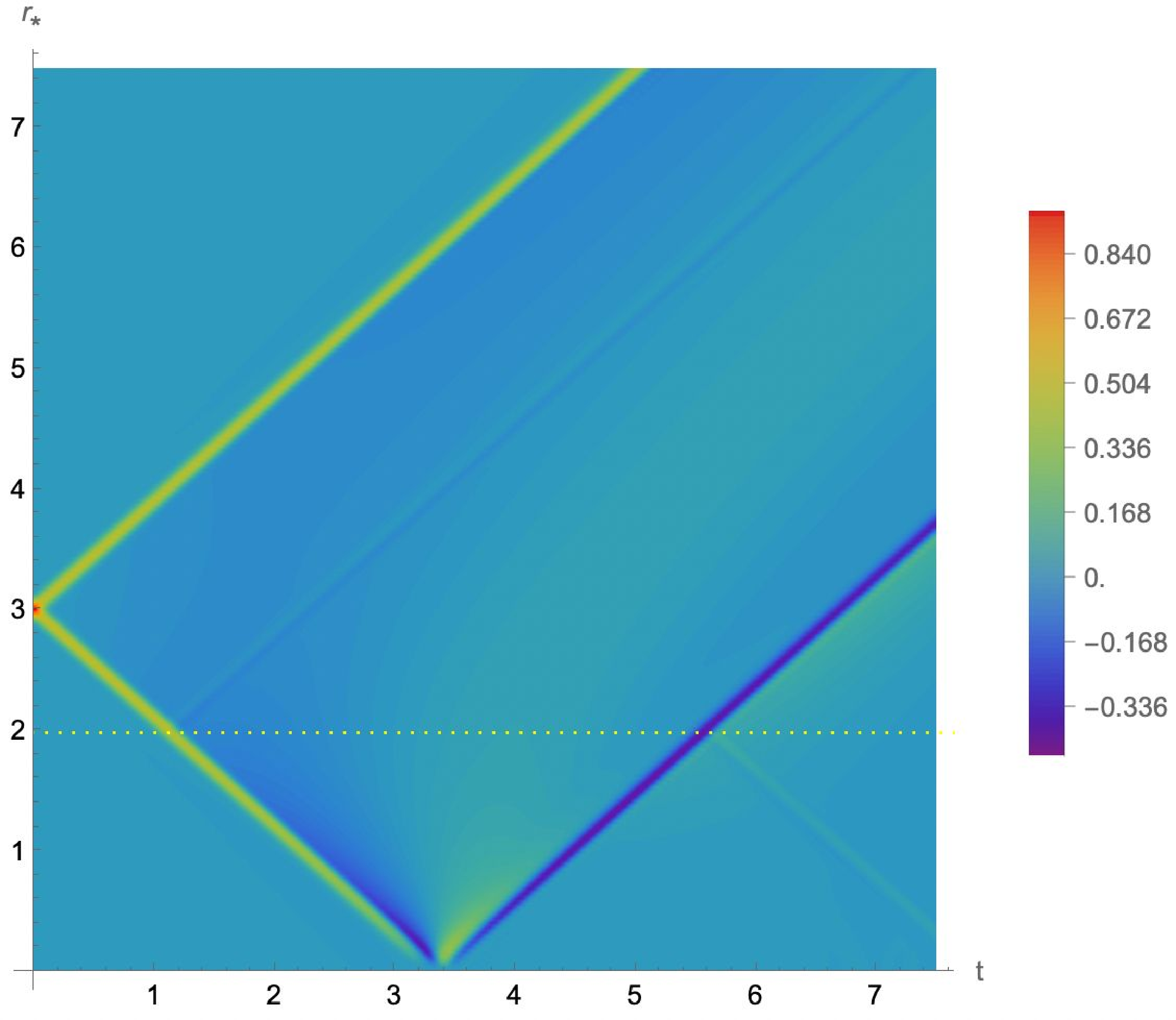}
\includegraphics[width=0.7\columnwidth]{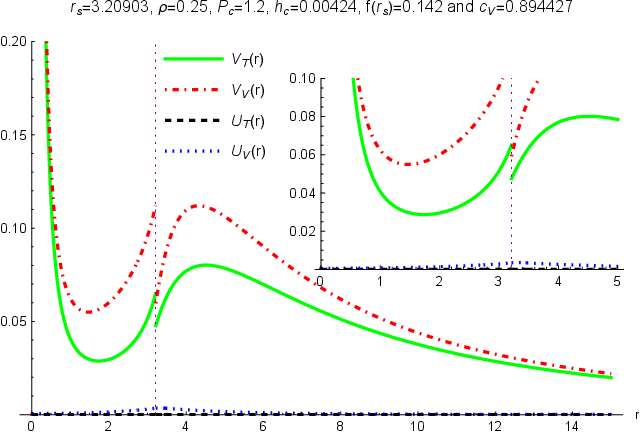}\includegraphics[width=0.7\columnwidth]{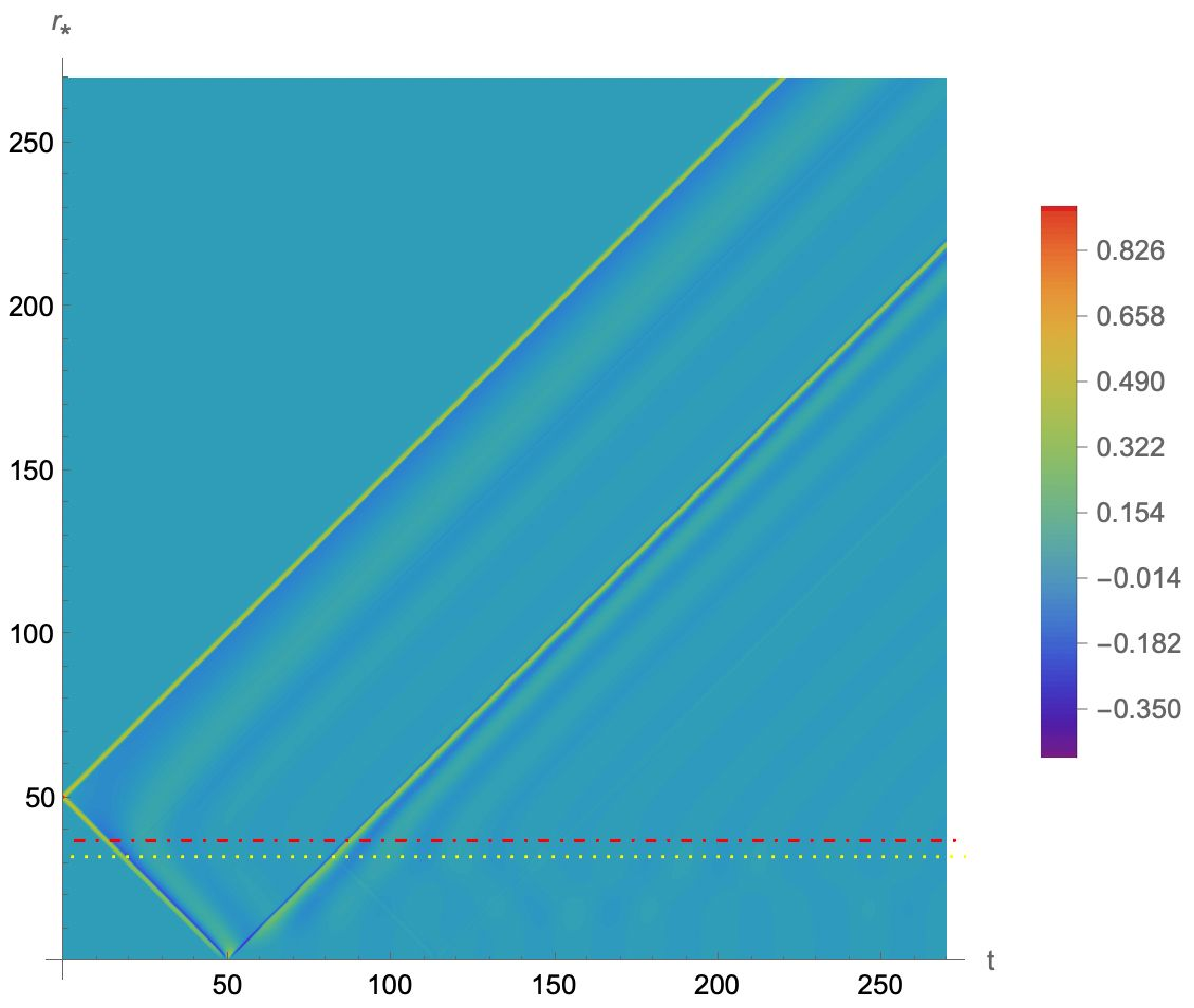}\includegraphics[width=0.7\columnwidth]{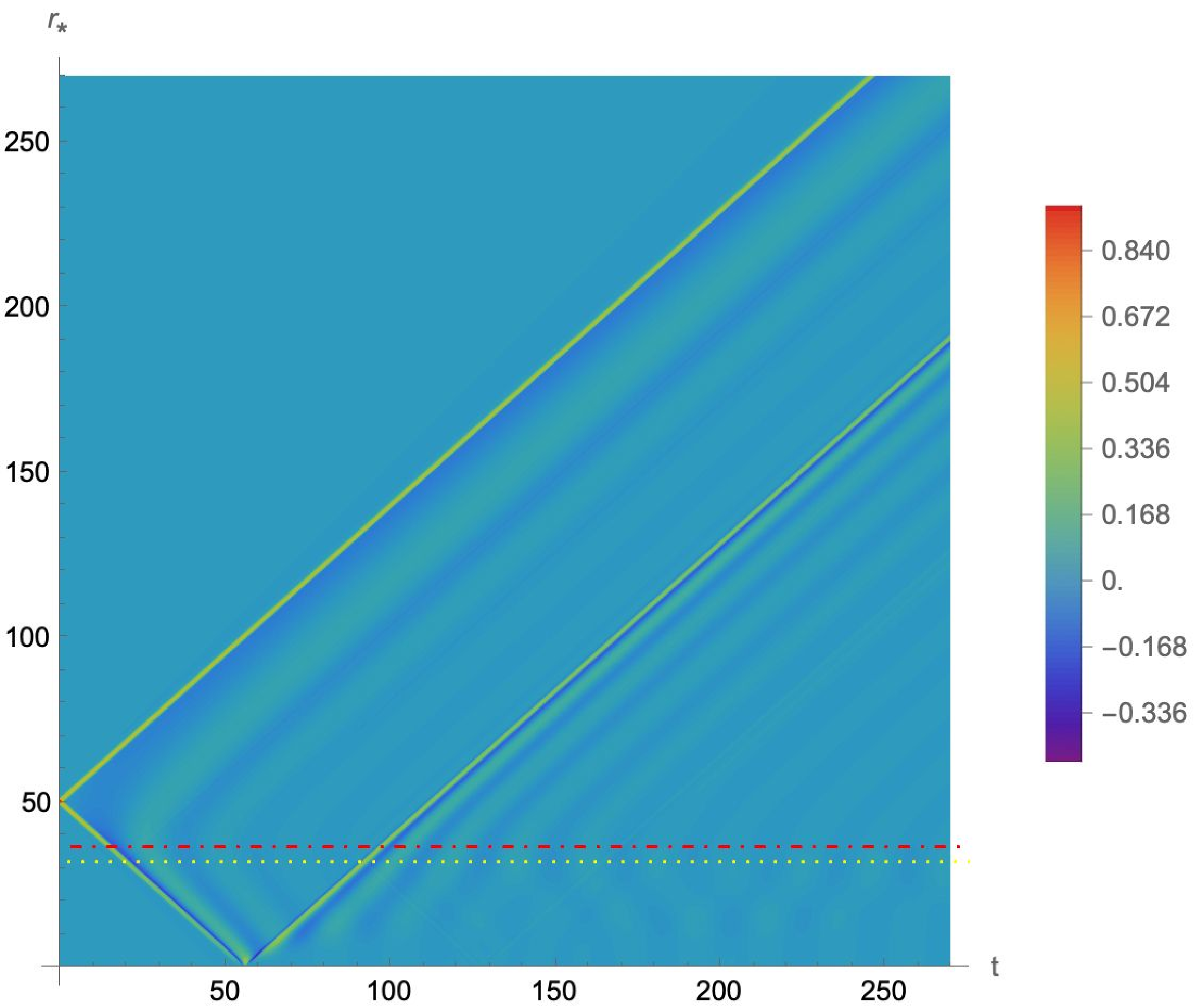}
\caption{Numerical results of the effective potentials (first column) and spacetime evolutions of the vector (mid column) and tensor modes (right column), associated with two distinct types (top and bottom rows) of echo mechanisms.
The calculations are carried out using the Einstein-{\AE}ther parameters Eqs.~\eqref{parLimGR}.
For the first type of effective potential shown in the top row, we consider $r_{s}=1.10657$, $\rho=1.2$, and $P_c=0.3$.
The second type presented in the bottom row utilizes the parameters $r_{s}=3.20903$, $\rho=0.25$, and $P_c=1.2$.
The spacetime evolutions are carried out for $R_B$ (mid column) and $R_C$ (right column) in a similar fashion shown in Fig.~4.
}
\lb{FigAppD2}
\end{figure*}

\begin{figure*}[htbp]
\centering
\includegraphics[width=0.85\columnwidth]{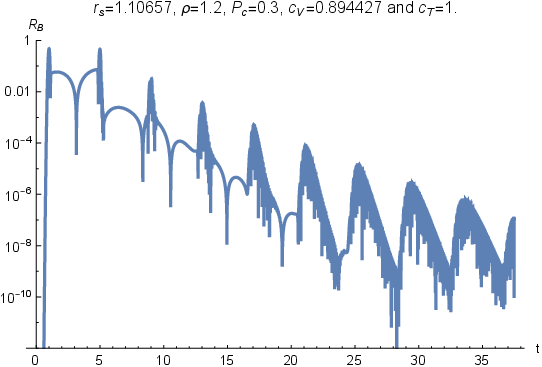}\includegraphics[width=0.85\columnwidth]{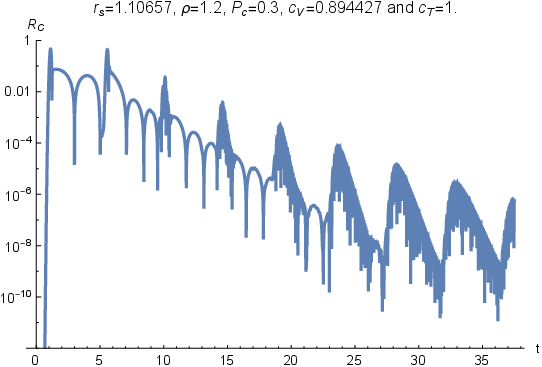}
\includegraphics[width=0.85\columnwidth]{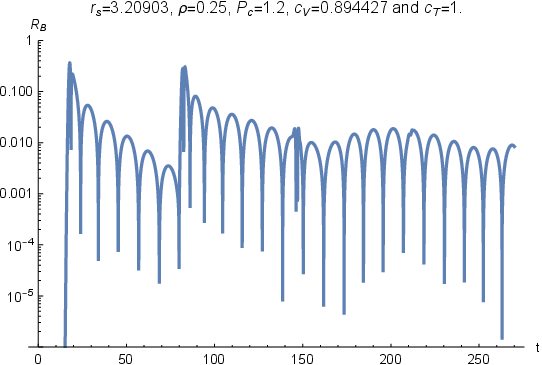}\includegraphics[width=0.85\columnwidth]{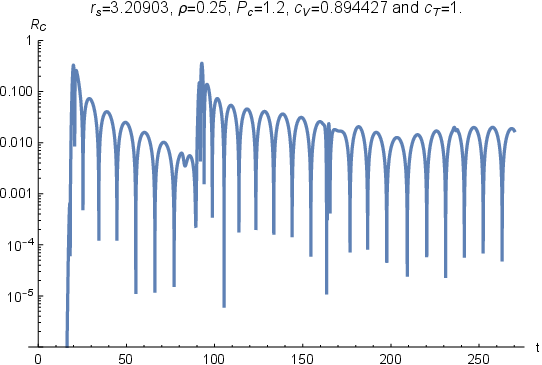}
\caption{Numerical results of temporal evolutions of the vector (mid column) and tensor modes (right column), associated with two distinct types (top and bottom rows) of effective potentials shown in Fig.~\ref{FigAppD2}, for which the same metric parameters are adopted.
The temporal evolutions of $|R_B|$ (left column) and $|R_C|$ (right column) are evaluated at $r=r_s$.
}
\lb{FigAppD3}
\end{figure*}

\section{Concluding remarks}
\renewcommand{\theequation}{4.\arabic{equation}} \setcounter{equation}{0}

This paper investigates the axial gravitational perturbations of a uniform density star in Scalar-Einstein-{\AE}ther gravity. 
The obtained results support the conclusions in~\cite{echoTwo2}, which indicate two echo types. 
One type originates from the existence of a potential well, and the other arises from the discontinuity at the star's surface.
As discontinuity is mainly present for most compact objects, the former type of echo, when it is relevant, is always mixed with the second type.
Also, the potentials involved in forming a potential well might have an irregular shape.
This makes it somewhat ambiguous to define the echo period rigorously.
However, this type of echo persists for an extensive period, and it might be relatively straightforward to distinguish the signals from the star's quasinormal oscillations.
In contrast, the latter type of echo might possess an unambiguous signal but might be short-lived.
This is because the system in question is more dissipative, as the star's surface generally does not efficiently confine the total energy.

We have also studied the properties of axial gravitational perturbations in a bimodal system, which possesses two distinct wave speeds, $c_V$ and $c_T$.
It is speculated that, owing to the coupling between these two modes, a shock wave might be formed.
However, numerical simulations indicate that such a phenomenon, if any, is significantly suppressed and does not lead to any observable outcome.
Nonetheless, the trajectories of wave propagation at two distinct speeds can be identified from the numerical calculations, which might not vanish in the general relativity limit.
Unfortunately, the potential distinct properties of the vector degree of freedom can hardly be utilized to extract or constrain relevant model parameters of the Einstein-{\AE}ther gravity, owing to the smallness of the couplings.
We plan to continue this line of research in future studies.

\begin{acknowledgments}
We are deeply grateful to the anonymous referee for their insightful comments, constructive critiques, and valuable suggestions. 
In particular, the discussions presented in Appx.~A have benefited significantly from their input.
This work is supported by the Brazilian agencies Fapesq-PB, FAPESP, FAPERJ, CNPq, and CAPES.
ARQ acknowledges CNPq support under process number 310533/2022-8.
\end{acknowledgments}

\appendix

\section{Israel junction conditions and the star metric with a vanishing scalar field}\label{appA}
\renewcommand{\theequation}{A.\arabic{equation}} \setcounter{equation}{0}

In this appendix, we show that the star metric ansatz Eqs.~\eqref{metric1} does not accommodate a nontrivial scalar ``hair''.
By using Eq.~\eqref{metric1}, one finds the following field equations:
\bqn
\lb{Equation1App}
&& h''+2\frac{h}{f}\frac{3P+\rho}{c_{14}-2}-\frac{h'^2}{h}+\frac{h'}{rf(c_{14}-2)}\left[r^2\rho\right.\nb\\
&&~~~~\left.+(2c_{14}-1)r^2P+(c_{14}-2)(f+1)\right] = 0 ,\nb\\
&&f'+f\left(\frac{2}{r}+\frac{h'}{h}\right)+\frac{2}{(2-c_{14})r}\left[r^2\rho\right.\nb\\
&&~~~~\left.+(2c_{14}-1)r^2P-2+c_{14}\right] = 0 ,\nb\\
&&\Psi'^2+\frac{f^{-1}-1}{r^2}+\frac{P}{f}-\frac{h'}{rh}-\frac{c_{14}h'^2}{8h^2} = 0 .
\eqn
For the exterior of the star $r>r_s$, we have $\rho=P=0$, and the vacuum solution reads:
\bqn
\lb{VacuumSolutionE}
f(r)&=&h(r)=1-\frac{2M}{r} ,\nb\\
\Psi(r)&=&\Psi_0+\frac{\sqrt{c_{14}}}{2\sqrt{2}}\ln{\frac{r}{r-2M}} ,
\eqn
for a non-vanishing scalar field, where $\Psi_0$ is a constant of integration.
The interior solution and the vacuum solution are connected at the star's surface by Israel's junction conditions: 
\bqn
f_\text{inside}(r_s) &=& f_\text{outside}(r_s), \nb\\
h_\text{inside}(r_s) &=& h_\text{outside}(r_s), \nb\\
{h'}_\text{inside}(r_s) &=& {h'}_\text{outside}(r_s), \nb\\
\Psi_\text{inside}(r_s)&=&\Psi_\text{outside}(r_s) , \label{bdConApp}
\eqn
together with a vanishing pressure $P(r_s)=0$, which is a generalization of Eq.~\eqref{bdCon}.

A few comments are in order.
The Israel junction condition dictates the mathematical feasibility when one glues two valid solutions of the Einstein field equations together via a hypersurface.
It was initially proposed for metric solutions in the vacuum~\cite{agr-collapse-thin-shell-03, book-general-relativity-Poisson}.
In the presence of a scalar field~\cite{agr-collspse-thin-shell-31}, the second junction condition is generalized to include a relation between the discontinuity of the first-order derivative of the scalar field and that of extrinsic curvature.
Notably, for the present case, where the scalar field is minimally coupled to the gravitational sector, it implies that the jump in the first-order derivative of the scalar field vanishes.
However, the third line of Eqs.~\eqref{Equation1App} indicates that the scalar field satisfies a first-order ordinary differential equation, and apparently, its boundary condition consists only of the first junction condition, which demands the field to be continuous on the surface.
In other words, the continuity condition of its first-order radial derivative, required by the second junction condition, cannot be straightforwardly enforced as a part of the boundary condition on the surface. 
In this regard, one seems to encounter the following dilemma: enforcing the second junction condition might lead to a trivial, namely, vanishing scalar field.
However, such difficulty is naturally alleviated by noting that the third line of Eqs.~\eqref{Equation1App} only involves $f$, $P$, $h$, and $h'$, which are manifestly continuous across the surface of the star.
In other words, the second junction condition for the scalar field, namely
\bqn
{\Psi'}_\text{inside}(r_s) = {\Psi'}_\text{outside}(r_s) ,
\eqn
is naturally derived from the third line of Eqs.~\eqref{Equation1App}.
Subsequently, in theory, the numerical solutions can be obtained given the boundary condition appropriately adapted to the Israel junction condition.
Specifically, one integrates numerically the field equations~\eqref{Equation1App} from the star's surface toward its center, with given $h(r_s), h'(r_s), f(r_s), \Psi(r_s)$ regarding the boundary conditions Eqs.~\eqref{bdConApp}.

However, it can be shown, from a rather general perspective, for the spherically symmetric case, the theory does not accommodate a nontrivial scalar ``hair'', minimally coupled to the gravitational sector, owing to the regularity condition of the scalar field.
To be specific, the latter implies that as $r \rightarrow 0$, the radial derivative of the scalar field must satisfy $\partial_{r}\Psi \rightarrow 0$. 
The scalar field obeys the Klein-Gordon equation, $\nabla_{\mu}\nabla^{\mu}\Psi = 0$, which reduces to
\bqn
\frac{1}{\sqrt{- g}}\partial_{r}\left( \sqrt{-g}g^{rr}\partial_{r}\Psi \right) = 0 ,
\eqn
by assuming spherical symmetry.
In the interior region where $r > 0$, both $\sqrt{-g}$ and $g^{rr}$ must remain finite to ensure the metric is invertible. 
Therefore, the Klein-Gordon equation implies that $\sqrt{-g}g^{rr}\partial_{r}\Psi$ must be constant throughout the interior. 
Taking the limit $r \rightarrow 0$, this constant is found to vanish, so we have
\bqn
\sqrt{-g}g^{rr}\partial_{r}\Psi = 0 \quad \text{in the interior}.
\eqn
Because both $\sqrt{-g}g^{rr}$ and $\partial_{r}\Psi$ must be continuous at the surface of the star, it follows that $\sqrt{-g}g^{rr}\partial_{r}\Psi = 0$ also holds in the exterior region. 
Again, since $\sqrt{-g}$ and $ g^{rr} $ are nonzero outside the star, this condition enforces $ \partial_{r}\Psi = 0 $ in the exterior as well. 
Therefore, one concludes that $\partial_{r}\Psi = 0$ everywhere, which is confirmed by explicit numerical integration.
By suppressing the scalar field, one falls back to a system of first-order field equations given by~\eqref{Equation1}, for which only the first Israel junction conditions are involved.

\section{The specific forms of the effective potentials for axial gravitational perturbations}\label{appB}
\renewcommand{\theequation}{B.\arabic{equation}} \setcounter{equation}{0}

In Eq.~(\ref{masterequation1}), the effective potentials are given by
\bqn
\lb{masterequation2}
V_T(r)&=&\frac{h}{r^2}\left(L^2+L-3+3f\right)\nb\\
&&+h\frac{\rho+(2c_{14}-1)P}{2-c_{14}}\nb\\
U_T(r)&=&\frac{2\sqrt{h}}{rBf}\left(\sqrt{2}\sigma_f-2f\right)\nb\\
V_V(r)&=&\frac{(10c_{14}^2-2+4c_{14}^2c_T^2-4c_V^2)Ph}{c_{14}c_V^2(c_{14}-2)}\nb\\
&&-\frac{c_{14}(c_V^2+13c_T^2-1)Ph}{c_{14}c_V^2(c_{14}-2)}\nb\\
&&+\frac{2c_Y^2-2-c_{14}(c_V^2+c_T^2-1)}{c_{14}c_V^2(c_{14}-2)}h\rho\nb\\
&&-2\sqrt{2}\frac{(c_{14}+2)c_V^2+(3c_{14}-4)c_T^2}{c_{14}^2c_V^2r^2}h\sigma_f\nb\\
&&+\frac{h}{c_{14}^2c_V^2r^2}\left\{c_{14}[4(c_{14}-1)c_T^2\right.\nb\\
&&+[2+c_{14}L(L+1)]c_V^2]+2f[(4\nb\\
&&+c_{14})c_V^2-2(2-c_{14})^2c_T^2+c_{14}(c_V^2\nb\\
&&\left.+2(c_{14}-1)c_T^2)r^2\Psi']\right\}\nb\\
U_V(r)&=&\frac{c_T^2h^{3/2}B}{c_{14}c_V^2r^3}(L^2+L-2)\left(\sqrt{2}\sigma_f-2f\right)
\eqn
where
\bqn
\lb{sigmaandB}
\sigma_f&=&\sqrt{f}\sqrt{f[2-c_{14}(r^2\Psi'^2-2)]+c_{14}(r^2\rho+1)}\nb\\
B'&=&\left[\frac{2}{r}-\frac{1}{rf}-\frac{(2c_{14}-1)P+\rho}{c_{14}-2}\frac{r}{f}\right]B
\eqn
The above potential functions are very complex.
However, if we take $c_i=0$, then $c_{14}=0$ and $c_T=c=1$, at which point $U_T=0$, the two master equations decouple, and one of them falls back to that governs the axial gravitational perturbations in general relativity.

\section{The feasibility of observing the vector degree of freedom in the present framework}\label{appC}
\renewcommand{\theequation}{C.\arabic{equation}} \setcounter{equation}{0}

Besides the specific model discussed in the present study, one can show schematically that as long as the coupling between the vector field and the gravitational sector is small, the bimodal effect elaborated in this manuscript can hardly lead to any observational implications.
On the one hand, as discussed, the smallness of the coupling, in terms of the coefficients $c_{i}$s, is attributed to the constraints associated with the (lack of) Cherenkov radiation~\cite{Elliott:2005va}.
On the other hand, as detectors are constituted by ordinary matter which, in the present framework, does not directly couple to the vector field, the strength of such an indirect interaction is dictated by the coefficients $c_i$s via the curved spacetime.
By including more matter fields, one may generalize Eq.~\eqref{Action1} to the following schematic form
\begin{equation}
S_{\text{tot }} = S_{\text{rest }}\left\lbrack \psi_{i}, g_{\mu\nu} \right\rbrack + \int
d^{4}x\sqrt{- g}\left( \lambda\left( u^{2} + 1 \right) + \varepsilon K(\nabla
u)^{2} \right) 
\end{equation}
where $\psi_{i}$ represents the matter fields embedded in their action $S_{\text{rest}}$. 
The operator $K$ symbolically indicates how the two factors of \(\nabla u\) are contracted.
Specifically, in the case of Eq.~\eqref{Action1}, we have $K \sim c_{1}g^{\alpha\beta}g_{\mu\nu} + \cdots$. 
The overall scale of the coefficients $c_i$s is denoted by $\varepsilon$, which we take to be small.

Formally, one can proceed to derive the field equations.
It is observed that the Lagrange multiplier scales as $\lambda\propto \varepsilon$ (cf. Eq.~\eqref{FieldEqu4}), which is obtained by contracting $u$ with the vector field equation.
This implies that the dynamics of the vector field is essentially governed by the relative magnitudes among the coefficients $c_i$s, as the overall scale $\varepsilon$ does not appear in the resulting field equation.
The vector degree of freedom might exhibit nontrivial behavior, such as possessing a different sound speed.
However, as one takes the limit $\varepsilon\to 0$, the field equations for the matter and metric fields reduce to those of general relativity.
This is because the contribution from the vector field is suppressed by a factor of $\varepsilon$, which is readily verified and already demonstrated by taking such a limit in Eqs.~\eqref{FieldEqu1} for the scalar and metric degrees of freedom.

As a result, the features associated with the vector degree of freedom can hardly lead to any observational effects.
We are indebted to the anonymous referee for pointing out the misunderstanding in the original version of the manuscript while providing a comprehensive explanation, which clarifies the potential relevance of the present study.

\end{document}